\documentclass[aps,physrev,twocolumn,groupedaddress]{revtex4-2}

\usepackage{graphicx}
\usepackage{epsf}
\usepackage{bm}
\usepackage{amsmath}
\usepackage{amsfonts}
\usepackage{amssymb}
\usepackage{epstopdf}
\usepackage{natbib}
\usepackage{nameref}
\usepackage{color}
\usepackage{verbatim}
\usepackage{multirow}
\usepackage{bm}
\usepackage{dcolumn}
\usepackage{placeins}

\usepackage{subfigure}
\definecolor{darkblue}{rgb}{0.0, 0.0, 0.55}
\definecolor{darkred}{rgb}{0.55, 0.0, 0.0}
\usepackage{hyperref}
\hypersetup{
    colorlinks=true, 
    linkcolor=darkblue,
    citecolor=darkblue,
    urlcolor=darkblue}
    
\makeatletter\let\expandableinput\@@input\makeatother

\DeclareTextCommand{\k}{OT1}[1]{{\fontencoding{T1}\selectfont\k{#1}}}

\begin{document}


\title{Exploring the dependence of giant radio source formation on the gas environment in dark matter halos}
\author{Xiaodong Duan}
\email[Contact author: ]{duanxiaodong@htu.edu.cn}
\affiliation{School of Physics, Henan Normal University, Xinxiang 453007, People’s Republic of China\\
Center for Theoretical Physics, Henan Normal University, Xinxiang 453007, People’s Republic of China.}


\date{\today}

\begin{abstract}
The known population of Mega-parsec giant radio sources (GRSs) including giant radio galaxies and quasars has soared from several hundred to more than $10^4$ in recent years. However, the formation mechanisms of GRSs remain elusive, and one explanation suggested is that they form in a low-density environment. 
In this work, we use magnetohydrodynamic simulations to study the formation of GRSs from hosts residing in dark matter halos with masses of  $10^{13}$, $10^{14}$ and $10^{15}$ solar masses, adopting different hot baryonic gas fractions (0.02--0.15) and varying density profiles. We inject jets into the host galaxies with an energy of 0.06 percent of the central black hole's relativistic energy and a power of 0.05 percent of the Eddington luminosity in most runs. 
GRSs can form in dark matter halos with normal hot baryonic gas fractions; however, the propagation of the radio lobes depends non-monotonically on the gas environment. The propagation of radio lobes can be slower in halos with sufficiently low or high central density and pressure, as a much lower central pressure cannot sufficiently collimate the jet and produces wider, less penetrating lobes, whereas an atmosphere with sufficiently high pressure enhances the interaction between the jet and the surrounding medium. Assuming equipartition between non-thermal electron and magnetic energy, the evolution of the simulated GRSs in the radio power--linear size diagram shows that the radio power of most simulated sources within halo masses of $\rm 10^{13}$ and $\rm 10^{14} M_\odot$ can reach values comparable to observational data at similar physical scales. The simulated sources with the same power but a shorter jet duration (54.8 Myr) become faint remnant sources when they propagate to GRS scales. 
\end{abstract}


\maketitle


\section{Introduction}
\label{section1}

Giant radio sources (GRSs), defined as those exceeding 0.7 Mpc in size and including giant radio galaxies (GRGs) and quasars (GRQs), have been known for a long time \citep{willis74, dabhade23}. The number of known GRSs has soared from several hundred to more than $10^4$, and the size record has been updated to approximately 7 Mpc in recent years  \citep{dabhade23, mostert24, oei24a, lao26}. The feedback from GRSs may be important for galaxy evolution \citep{sorini22, cen24, qiu25}. However, the formation mechanism of these mega-parsec GRSs remains unclear.

Several mechanisms \citep{dabhade23}, including a low-density environment, unusually high jet power \citep{Gopal-Krishna89, Wiita89}, and recurrent jet activity, have been proposed to explain the formation of GRSs.  The low-density environment is the most widely invoked explanation for the formation of GRSs \citep{chen11, dabhade20a, lan21, oei24b}. By adopting the optical velocity dispersion of group members within 0.7 Mpc of GRG NGC 6251, \cite{chen11} indirectly suggested that the density of X-ray emitting gas in NGC 6251 is unusually low. The most recent observations show that GRSs can form in nearly all cosmic environments: about 30\% of all luminous GRSs reside in galaxy clusters, 60\% in galaxy groups, and roughly 10\% in more dilute regions such as filaments, sheets, or voids \citep{oei24a, oei24b}. However, these observations have not directly related GRSs to the local gas environment, in particular the baryonic gas fractions within the virial radius of dark matter halos, through which the jets propagate, while strong jet–ambient gas interactions will suppress the growth of large-scale lobes \citep{duan25}. Besides the interstellar medium (ISM), the local gas environment of the host halos primarily consists of hot circumgalactic medium (CGM) within the dark matter halos for hosts that are massive galaxies (\citealt{tumlinson17}), and of intragroup or intracluster medium (ICM) for hosts that are central giant elliptical galaxies in the galaxy groups or clusters (\citealt{fabian12}). The massive satellites could also retain hot CGM \citep{rohr24, zhang24}. The amount of hot gas in a dark matter halo is characterized by its hot baryonic gas fraction, which for low-redshift galaxies primarily depends on the mass of dark matter halo \citep{xu20, dev24}. Future observations may probe the relation between the formation of GRSs and the local gas environment, particularly in dark matter halos of different masses and baryonic gas fractions (e.g., LOFAR, SKA and Hongmeng \citep{chen23-hongmeng, an26-hongmeng} combined with eROSITA).   

Active galactic nuclei (AGN) jet feedback, particularly in cluster environments, has been widely studied with hydrodynamic (HD) and magnetohydrodynamic (MHD) numerical simulations \citep{omma04, gaspari11, li15, weinberger17, weinberger23, guo18, duan20, ehlert23, dominguez-Fernandez24}. The evolution of radio jets on galactic scales is also extensively explored in simulations, especially regarding the morphological distinction between Fanaroff--Riley type I and II sources \citep{tchekhovskoy16, massaglia16, massaglia22, perucho19, perucho23, yates23}. However, simulations addressing the formation of GRSs remain rare (no MHD simulation of a GRS existed until 2024, as noted by \citet{oei24a}). Recently, MHD simulations have been applied to study the formation of GRSs \citep{duan25, giri25a, giri25b}. Focusing on the formation of GRSs in cluster environments with virial masses of $\rm 10^{14} M_\odot$ including a central ISM and ICM setup, \citet{duan25} presented a study of key jet parameters---such as power, energy composition, and magnetic field structure for the GRS jets---using MHD simulations. We briefly summarize the results as follows: 

(i) The most significant parameter is jet power, which exhibits a ``lower power--larger bubble'' effect: for a fixed total jet energy ($\rm 2.06\times10^{59}\ erg$) with different jet durations, lower-power jets with longer durations tend to produce larger radio sources, while higher-power jets with shorter durations are injected with higher internal pressure, causing them to expand laterally and penetrate less deeply into the ambient medium. Additionally, higher-power jets dissipate their energy more rapidly in the early stages via adiabatic losses. 

(ii) The magnetic field structure also has a significant impact: strong poloidal fields hinder radio lobe growth, a result also reported by \cite{chen23}. The influence of energy composition is less pronounced than that of jet power or magnetic field structure. 

(iii) All simulated low-power jets ($\rm 10^{-4} - 10^{-3} L_{Edd}$) from low-spin black holes can produce GRS-scale ($> 0.7$ Mpc) sources, where $\rm L_{Edd}$ is the Eddington luminosity of the central black hole. Analysis of the power--linear size (P--D) diagram shows that jets with a power of $\rm \sim 5 \times 10^{-4} L_{Edd}$ (corresponding to a jet duration of 200 Myr in \cite{duan25}) and an initially sufficient toroidal magnetic field yield radio sources that best match current observations.

\citet{giri25a} and \citet{giri25b} show that Mpc scale radio sources  form in all of their simulations from hosts in galaxy groups with a halo mass of $\rm 10^{13} M_{\odot}$, which varied in jet power and the location of the jet base in a triaxial medium without a central ISM setup. This suggests that GRSs may be more common than previously believed. Furthermore, \citet{giri26a} and \citet{giri26b} show that the ISM gas does not significantly influence the evolution of jetted lobes produced by extremely powerful jets ($\rm 2.4 \times 10^{46}\ erg\ s^{-1}$) within a dark matter halo of $\rm 10^{13} M_{\odot}$, but it is important for the jet power in more general cases with lower powers. On the other hand, all these simulations have not focused on the formation of GRSs in different local gas environments with different baryonic gas fractions. However, varying hot baryonic gas fractions are indicated by observations, particularly for halos of $10^{13}$ and $10^{14}$ solar masses \citep{dev24}. In this study, we use numerical simulations to investigate the formation and properties of GRSs within dark matter halos of varying masses ($\rm 10^{13}$--$\rm 10^{15} M_{\odot}$) and hot baryonic gas fractions (0.02--0.15).


This paper is structured as follows. In Sect. \ref{section2}, we describe the basic methodology. In Sect. \ref{section3}, we present an investigation of GRS formation in galaxies with varying masses, hot baryonic gas fractions, density profiles, and jet properties. Finally, in Sect. \ref{section4}, we summarize the key findings.

\section{Methods}
\label{section2}

In this study, we update the methods used in \cite{duan25} to investigate the formation of GRSs in dark matter halos with masses of $10^{13}$, $10^{14}$, and $10^{15}$ solar masses. The main updates include: (1) The concentration parameter of the dark matter halo is related to virial mass and redshift; (2) A core parameter $\rm a_{core}$ is introduced to describe the degree of gas concentration in the halo center; (3) The estimate of reasonable jet power incorporates more detailed black hole accretion physics; (4) When calculating radio emission, the energy density of non-thermal electrons is set assuming equipartition with the magnetic energy density; (5) To approximately account for the cooling of non-thermal electrons, we adopt a spectral index $\alpha$ that varies with radio lobe length. The main methods are summarized as follows.    

\subsection{Numerical setup}
\label{section2.1}

\begin{figure*} [!t]
\centering
\includegraphics[height=0.24\textheight]{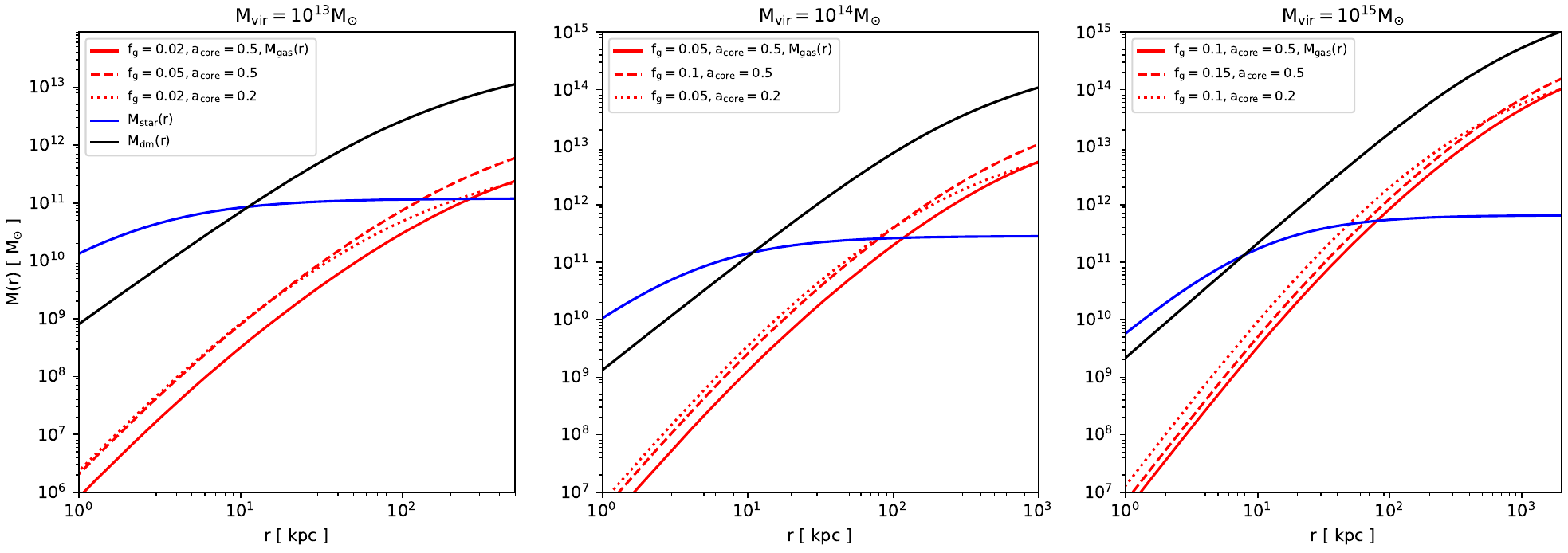}
\caption{ Profiles of enclosed mass within radius $r$ for different components: hot gas $\rm M_{gas}(r)$ (red lines), dark matter $\rm M_{dm}(r)$ (dark solid lines), and stars $\rm M_{star}$ (blue solid lines), for virial halo masses of $\rm 10^{13}$ (left), $\rm 10^{14}$ (middle), and $\rm 10^{15}$ (right) solar masses. Different line styles correspond to varying hot gas fractions ($\rm f_g$) and core parameters ($\rm a_{core}$).}
\label{figM-r}
\end{figure*}

We mainly solve the ideal MHD equations numerically in 2.5-dimensional cylindrical coordinates using MPI-AMRVAC 2.0 \citep{xia18},
\begin{equation}
\rm \frac{\partial \rho}{\partial t} + \nabla \cdot( \rho \textbf{v}) = \dot{\rho}_{\text{inj}}, 
\end{equation}
\begin{equation}
\rm \frac{\partial (\rho \textbf{v}) }{\partial t} + \nabla \cdot ( \rho \textbf{vv}  - \textbf{B} \textbf{B} ) +\nabla \Bigl(p + \frac{B^2}{2}\Bigr) + \rho \nabla \Phi = \dot{\rho}_{\text{inj}} \textbf{v}_{\text{inj} }, 
\end{equation}
\begin{equation}
\rm \frac{\partial e }{\partial t} + \nabla \cdot \Bigl[ \Bigl(e + p + \frac{B^2}{2}\Bigr) \textbf{v} - \textbf{B} \textbf{B} \cdot \textbf{v} \Bigr] + \rho \textbf{v} \cdot \nabla \Phi = \dot{e}_{\text{inj}}, \label{eq-e} 
\end{equation}
\begin{equation}
\rm \frac{\partial \textbf{B} }{\partial t} + \nabla \cdot  (\textbf{v}  \textbf{B} - \textbf{B} \textbf{v} ) = \dot{\textbf{B} }_{\text{inj}}. \label{eq-B}
\end{equation}
In these equations, $\rm \rho$, $\rm \textbf{v}$, and $\rm p$ are the density, velocity, and thermal pressure, respectively. $\rm e$ is the total energy density, including thermal, kinetic, and magnetic energy. $\rm \textbf{B}$ is the magnetic field in code units, convertible to Gauss units via $\rm \textbf{B} (\text{G}) = \sqrt{4\pi} \textbf{B}$. $\rm \Phi$ is the gravitational potential. The source terms on the right-hand sides represent jet injection while the jets are active. Radiative cooling, which is important for studying feedback processes \citep{duan24}, is not included in this work, as we focus primarily on the dynamic evolution of jetted bubbles. The equations are closed by the equation of state,
\begin{equation}  
\rm p = (\gamma_a - 1) \Bigl( e -  \frac{\rho v^2}{2} - \frac{B^2} {2} \Bigr),
\end{equation}
where $\rm \gamma_a = 5/3$. We simulate only one side of the bipolar jets. Stretched meshes are generally used in grid-based astrophysical fluid dynamics codes to simulate stratified atmospheres \citep{mignone07, springel10, guo14, guo18, xia18}. In this work, using the static adaptive mesh refinement in AMRVAC, the computational domain is initially refined with different spatial resolutions: 0.25 kpc within 100 kpc, 0.5 kpc from 100 to 200 kpc, 1 kpc from 200 to 400 kpc, and 2 kpc from 400 kpc to the outer boundary. The outer boundaries are set at 800 kpc, 1000 kpc, and 2000 kpc for dark matter halos with virial radii of 416 kpc, 896 kpc, and 1931 kpc, respectively. The resolution has been changed gradationally among regions refined with different levels of refinement in AMRVAC to avoid abrupt change of resolutions. We also perform a three-dimensional simulation in Cartesian coordinates for comparison and discuss it in Sect.~\ref{section3.3}. 

\begin{figure*} 
\centering
\includegraphics[height=0.47\textheight]{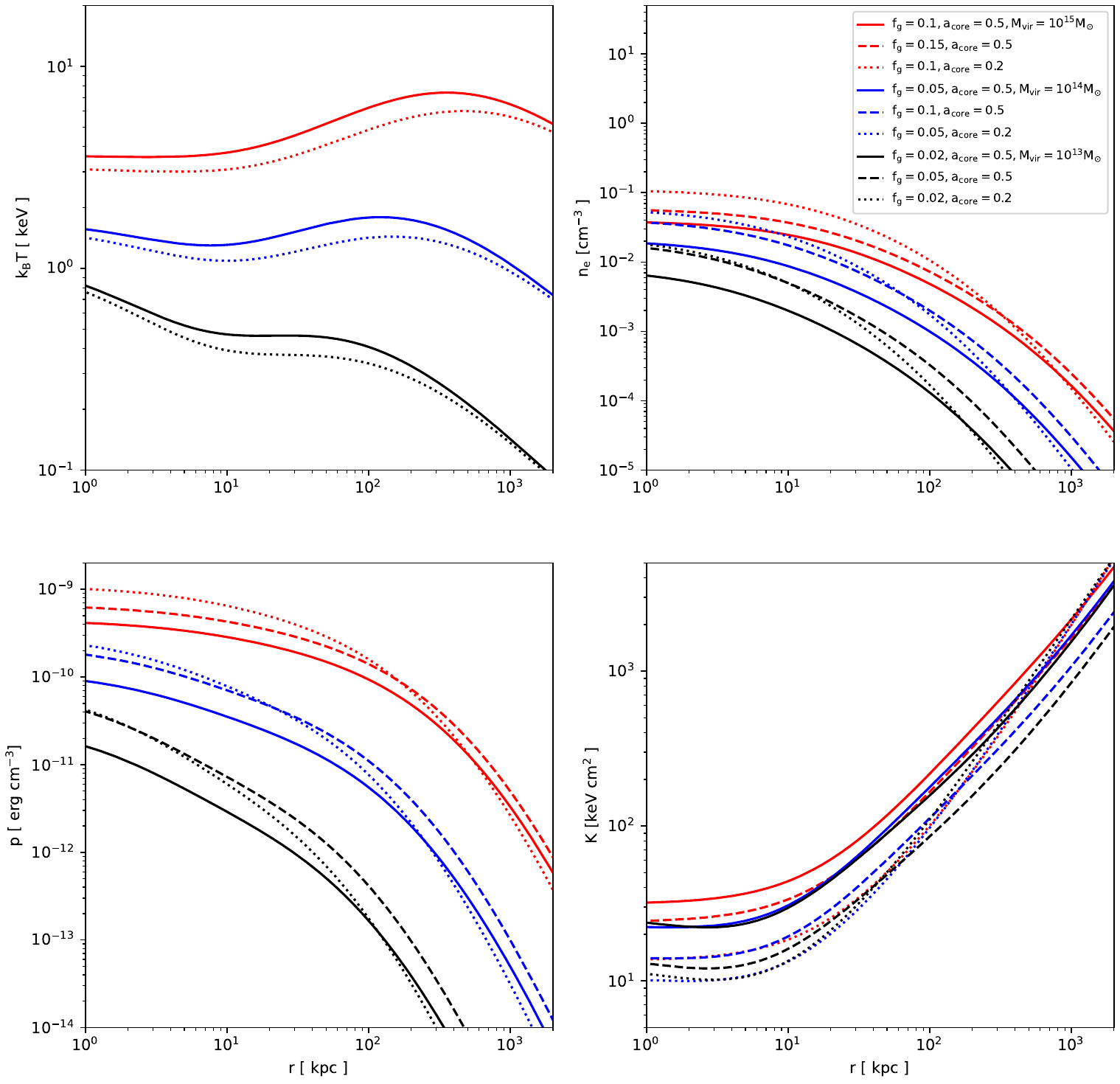}
\caption{Profiles of temperature $\rm T$, electron number density $\rm n_e$, pressure $\rm p$, and entropy index ($\rm K = k_BT/n_e^{2/3}$) of the hot gas for dark matter halos with virial masses of $10^{13}$ (black), $10^{14}$ (blue), and $10^{15}$ (red) solar masses. Different line styles correspond to varying hot gas fractions ($\rm f_g$) and core parameters ($\rm a_{core}$). In our model, the temperature profiles are independent of the gas fraction. Profiles outside the virial radius are fitted as extrapolations of the inner profiles. }
\label{figrhoTp}
\end{figure*}

\subsection{Galaxy model}
\label{section2.2}

We model galaxies within the framework of the current standard model of galaxy formation in a $\Lambda$CDM universe \citep{bookSchneider15}. The gravitational potential is mainly contributed by the dark matter halo, the stellar component of the central galaxy, and the central supermassive black hole, as adopted in previous studies \citep{guo18, duan18, duan20, fang20}. The dark matter gravitational potential follows the Navarro--Frenk--White (NFW) profile \citep{navarro97}:
\begin{equation}
\rm \Phi_{\text{dm}} = -\frac{2GM_0}{r_\text{s}} \frac{\ln (1+r/r_\text{s})}{r/r_\text{s}},
\end{equation}
where $\rm M_0$ and $\rm r_\text{s}$ are determined by the virial mass of the dark matter halo $\rm M_{\text{vir}}$ as
\begin{equation}
\begin{aligned}
\rm M_0 &= \frac{\rm M_{vir}/2} {\rm  \ln(1+C) -C/(1+C) }, 
\rm r_s &= \frac{R_{vir}}{C},
\end{aligned}
\end{equation}
with $\rm R_{vir}$ the virial radius given by
\begin{equation}
\rm R_{vir} = 206 \left(\frac{M_{vir}}{10^{12}M_{\odot}}\right)^{1/3} kpc.
\end{equation}
In this study, we adopt a concentration parameter that depends on virial mass and redshift,
\begin{equation}
\rm C \approx 6.7 \left(\frac{M_{\rm vir} }{ 2\times 10^{12} h^{-1} M_{\odot}}\right)^{-0.1} (1+z)^{-0.5},
\end{equation}
appropriate for NFW profiles in a standard $\Lambda$CDM universe \citep{duffy08, bookSchneider15}. We adopt a scaled Hubble constant $\rm h = 0.7$ and a redshift $\rm z = 0.2$ in this work.

The potential of the stellar component in the central galaxy \citep{hernquist90} is given by
\begin{equation}
\rm \Phi_{\star} = -\frac{GM_{\star}}{r+a},
\end{equation}
where $\rm M_{\star}$ is the total stellar mass, $\rm a = R_e/1.8153$, and $R_e$ is the radius of the isophote enclosing half of the galaxy's light. The stellar mass $\rm M_{\star}$ is determined by $\rm M_{\text{vir}}$ \citep{guoqi10} as
\begin{equation}
\rm M_{\star} = 0.129M_{vir} C_M^{-2.44},
\end{equation}
where the function $\rm C_{M}$ is
\begin{equation}
\rm C_{M} = \left(\frac{M_{vir}}{2.5119\times 10^{11} M_{\odot}}\right)^{-0.926} + \left(\frac{M_{vir}}{2.5119\times 10^{11} M_{\odot}}\right)^{0.261}.
\end{equation}
We fit $\rm R_e$ from \cite{jin20} using the relation
\begin{equation}
\rm R_e (kpc) = \left( \frac{0.186 M_{\star}}{10^{10} M_{\odot}} \right)^{1.116} + 1.17.
\end{equation}

The potential of the central black hole is adopted as a pseudo-Newtonian potential \citep{paczy80}:
\begin{equation}  
\rm \Phi_\text{bh} = -\frac{GM_\text{bh}}{r-r_g},
\end{equation}
where $\rm M_\text{bh}$ is the black hole mass and $\rm r_\text{g} = 2GM_\text{bh}/c^2$ is the Schwarzschild radius. The black hole mass is related to the stellar mass by \citep{haring04}
\begin{equation}  
\rm M_\text{bh} = 1.6 \times 10^8 \left( \frac{M_{\star}}{10^{11} M_{\odot}} \right)^{1.12} M_{\odot}.
\label{mbh}
\end{equation}

The gas profiles in the host halos of GRSs at the onset of activity have not been observed in detail so far, so we model them with more general profiles for the initial setup. We modify the model of \cite{fang20} and \cite{guo20b} to set up the hot gas atmosphere. The density profile of the hot gas is given by
\begin{equation}  
\rm \rho(r) = \frac{M_\text{n}}{(r+R_e)(r+a_{\rm core} R_{vir})^2},
\end{equation}
where $\rm a_{core}$ is the gas core parameter (smaller values correspond to a denser gas core); we mainly adopt $\rm a_{core} = 0.5$ and $0.2$ for parameter studies. $\rm M_\text{n}$ is a normalization constant related to the total gas mass $\rm M_\text{g}$ within the virial radius via
\begin{equation}  
\rm M_{g} = \int_{r_{min}}^{r_{vir}} 4\pi r^2 \rho(r) \, dr.
\end{equation}
The total gas mass is given by $\rm M_{g} = f_{g} M_{vir}$, where $\rm f_{g}$ is the hot baryonic gas fraction within the virial radius $R_{vir}$, which includes both the hot ISM of the central hosts and the intra-group or intra-cluster gas in dark matter halos of $10^{13}$, $10^{14}$, and $10^{15} M_\odot$. Following recent observations of \cite{dev24}, we adopt standard hot gas fractions $\rm f_g$ of 0.02, 0.05, and 0.1 for $\rm M_{vir} = 10^{13}$, $10^{14}$, and $10^{15}$ solar masses, respectively, in our fiducial runs. As different values of $\rm f_{g}$ are also suggested by observations, particularly for halos of $10^{13}$ and $10^{14}$ solar masses (see the comparison in \cite{dev24}), we adopt different values of $\rm f_g$ for parameter studies, as shown in Tab.~\ref{tab1}. 

Once the gravitational potential and density profile are determined, the initial temperature profile is set by assuming hydrostatic equilibrium, $\rm \rho \nabla \Phi = - \nabla p$, with $\rm p = \rho k_B T/(\mu m_p)$ and a mean molecular weight $\rm \mu = 0.61$. The temperature at the outer boundary is set to $\rm T_{out} = 0.5 T_{vir}$, where the virial temperature $\rm T_{vir}$ is
\begin{equation}  
\rm T_\text{vir} = 10^6 \left( \frac{M_\text{vir}}{ 10^{12} M_{\odot}} \right)^{2/3} K.
\end{equation}

The main input galaxy parameters are $\rm M_{vir}$, the hot gas fraction $\rm f_g$, and the gas core parameter $\rm a_{core}$. The values of $\rm f_g$ and $\rm a_{core}$ vary across simulations for parameter studies, as listed in Table \ref{tab1}. Profiles of the enclosed mass within radius $\rm r$ for different components---hot gas $\rm M_{gas}(r)$, dark matter $\rm M_{dm}(r)$, and stars $\rm M_{star}$---are shown in Fig. \ref{figM-r} for dark matter halos of different virial masses. The gas density, temperature, and pressure profiles are illustrated in Fig. \ref{figrhoTp}. These show that lower hot gas fractions result in markedly lower pressure profiles. The resulting gas profiles can be compared with those compiled in the literature for dark matter halos of comparable mass \citep{fabian12, mcNamara12, guo14, hogan17a, hogan17b, lakhchaura18, werner19}. The temperature and entropy profiles for halo masses of $\rm 10^{14}$ and $\rm 10^{15} M_{\odot}$ exhibit the typical characteristics of cool-core clusters, which are usually observed to be radio-loud or to host X-ray cavities.

\begin{table*}[t]
 \centering
\caption{The parameters for the galaxies and jets in our simulations. The parameters encompass jet energy $\rm E_{ jet}$, ratio of jet energy to relativistic energy of the black hole $\rm E_{jet} / E_{bh}$, jet power $\rm P_{ jet}$, ratio of jet power to Eddington luminosity $\rm P_{ jet} / L_{Edd}$, injected magnetic energy fraction $\rm f_{m}$, jet duration $\rm t_{jet}$, hot gas fraction $\rm f_{g}$ and gas core parameter $\rm a_{core}$ . The symbol “—” indicates that the corresponding parameter shares the same value as the fiducial runs ( M13fg02, M14fg05, and M15fg1). ``M13'', ``M14'' and ``M15'' in the run names denote virial masses of $\rm 10^{13}$, $\rm 10^{14}$, and $\rm 10^{15}$ solar masses, respectively.}
\label{tab1}
\begin{tabular}{lccccccccc}
\hline\hline 
Run & $\rm E_{ jet}(erg) $ & $ \rm E_{ jet} / M_{bh}c^2 $ &  $\rm P_{ jet}(erg\ s^{-1})$ & $\rm P_{ jet} / L_{Edd} $ & $\rm f_{m}$ & $\rm t_{jet} (Myr)$ & $\rm f_{g}$ & $\rm a_{core}$  \\
\hline
 M13fg02              &   $2.2 \times 10^{59}$ & $6.2 \times 10^{-4}$  & $ 1.27\times 10^{43}  $  & $5\times 10^{-4}$  & 0.8  &548  & 0.02  & 0.5  \\
 M13fg03      &   ---                                & ---                              & ---                                   & ---                           & ---   &---     & 0.03  & ---   \\
 M13fg04       &   ---                                & ---                              & ---                                   & ---                           & ---   &---     & 0.04  & ---   \\
 M13fg05      &   ---                                & ---                              & ---                                   & ---                           & ---   &---     & 0.05  & ---   \\
 M13fg1     &   ---                                & ---                              & ---                                   & ---                           & ---   &---     & 0.1  & ---   \\
 M13fg02core2   &  ---                                & ---                               & ---                                   & ---                           & ---   &---     & ---    & 0.2  \\
 M13fg02fk9       &   ---                               & ---                                & ---                                   & ---                         & 0.05 &---     & 0.02     & ---  \\
 M13fg03fk9       &   ---                               & ---                                & ---                                   & ---                         & 0.05 &---     & 0.03     & ---  \\
 M13fg04fk9       &   ---                               & ---                                & ---                                   & ---                         & 0.05 &---     & 0.04     & ---  \\
 M13fg05fk9       &   ---                               & ---                                & ---                                   & ---                         & 0.05 &---     & 0.05     & ---  \\
 M13fg1fk9       &   ---                               & ---                                & ---                                   & ---                         & 0.05&---     & 0.1     & ---  \\
 M13fg02t0E0    &  $2.2 \times 10^{58}$   & $6.2 \times 10^{-5}$   &   ---                                 & ---                          &---   & 54.8  & ---     & ---    \\
\hline
 M14fg05             &   $5.8 \times 10^{59} $ &  $6.2 \times 10^{-4}$ &  $ 3.35\times 10^{43}  $  &$5\times 10^{-4}$   & 0.8   &548  & 0.05  & 0.5   \\
 M14fg02       &   ---                                &   ---                           &  ---                                     &---                           & ---   &---      & 0.02   & ---  \\
 M14fg03       &   ---                                &   ---                           &  ---                                     &---                           & ---   &---      & 0.03   & ---  \\
 M14fg04       &   ---                                &   ---                           &  ---                                     &---                           & ---   &---      & 0.04   & ---  \\
 M14fg1       &   ---                                &   ---                           &  ---                                     &---                           & ---   &---      & 0.1   & ---  \\
 M14fg1-3D       &   ---                                &   ---                           &  ---                                     &---                           & ---   &---      & 0.1   & ---  \\
 M14fg05core2   &   ---                                &   ---                           &  ---                                     &---                           & ---   &---      & ---    & 0.2 \\
 M14fg02fk9       &   ---                                &   ---                           &  ---                                     &---                           & 0.05 &---      & 0.02  & ---  \\
 M14fg03fk9       &   ---                                &   ---                           &  ---                                     &---                           & 0.05 &---      & 0.03  & ---  \\
 M14fg04fk9       &   ---                                &   ---                           &  ---                                     &---                           & 0.05 &---      & 0.04  & ---  \\
 M14fg05fk9       &   ---                                &   ---                           &  ---                                     &---                           & 0.05 &---      & 0.05  & ---  \\
 M14fg1fk9       &   ---                                &   ---                           &  ---                                     &---                           & 0.05&---      & 0.1  & ---  \\
 M14fg05t0E0    & $5.8 \times 10^{58} $    & $6.2 \times 10^{-5}$ &  ---                                     &---                           & ---   & 54.8    & ---    & --- \\
\hline
 M15fg1           &  $1.5 \times 10^{60} $   & $6.2 \times 10^{-4}$  &  $ 8.57\times 10^{43}  $  &  $5\times 10^{-4}$   &0.8   &548    & 0.1 & 0.5   \\ 
 M15fg15       &  ---                               &  ---                              &  ---                                    & ---                          & ---   &---      & 0.15  & ---  \\ 
 M15fg1core2   &    ---                             &  ---                              &  ---                                   & ---                          & ---   &---       & ---      & 0.2   \\ 
 M15fg1fk9       &   ---                             &  ---                               &   ---                                  & ---                          & 0.05 &---      & ---      & ---  \\ 
 M15fg1t0E0    & $1.5 \times 10^{59} $  & $6.2 \times 10^{-5}$    &  ---                                   & ---                          & ---   & 54.8   & ---      & ---   \\ 
\hline
\end{tabular}
\end{table*}

\subsection{Jet injection}
\label{section2.3}

In this work, we initialize the jets in all runs with a varying magnetic energy fraction $\rm f_m$ and a fixed thermal energy fraction $\rm f_{th} = 0.05$, such that the kinetic energy fraction is $\rm f_{k} = 1 - f_m - f_{th}$. The jets are activated at the start of the simulation and maintain constant power during the active phase ($\rm t \leq t_{jet}$), where $\rm t_{jet}$ is determined by the adopted jet energy and power. Throughout this phase, mass, momentum, and non-magnetic energy are injected into a cylindrical region at the center of the gas atmosphere \citep{omma04, li15, duan20}. This region has a cross-sectional radius of $\rm R_{0} = 1$ kpc and extends 1 kpc along the $\rm +z$ axis. The jet velocity is fixed at $\rm 0.1c$. We inject the magnetic field using the magnetic-energy-controlled method detailed in \cite{duan25}, with a magnetic structure that follows \cite{gan17} and \cite{mbarek19} with modifications. The toroidal magnetic field is set proportional to $\rm r$ inside the jet base, to $\rm 1/r$ outside, and to zero farther out, as given by
\begin{equation}  
\rm B_{{inj},\ \phi} =
\left \{
\begin{aligned}
\rm B_0 \frac{r}{R_0},\  \ & \text{for}\   \rm r < R_0 \text{ and } z < h_{\rm jet}, \\
\rm B_0 \frac{R_0}{r},\  \ & \text{for}\   \rm R_0 < r < 2R_0 \text{ and } z < h_{\rm jet}, \\
\rm 0,\  \ & \text{for}\  r > 2R_0\  \text{or}\   \rm z > h_{\rm jet},
\end{aligned}
\right .
\label{bphi}
\end{equation}
where $\rm R_0$ and $\rm h_{\rm jet}$ are the cross-sectional radius and height of the jet base, respectively.
The poloidal magnetic field components are defined as
\begin{equation}  
\rm B_{\text{inj},\ z} = \alpha_p B_0 \cdot 2\left(1 - \frac{r^{2}}{R_{0}^{2}}\right) \exp\left(-\frac{r^2+z^2}{R_{0}^{2}}\right), 
\label{bz}
\end{equation}
\begin{equation}  
\rm B_{\text{inj},\ r} = \alpha_p B_0 \cdot \frac{2zr}{R_0^2} \exp\left(-\frac{r^2+z^2}{R_{0}^{2}}\right),
\label{br}
\end{equation}
where $\rm \alpha_p$ determines the ratio between the poloidal and toroidal magnetic fields. For all runs in this work, $\rm \alpha_p = 0.3$, indicating that the injected magnetic field is dominated by the toroidal component. The poloidal field is injected throughout the entire simulation domain, though its value is negligible far from the jet base. This magnetic field configuration analytically satisfies $\rm \nabla \cdot \textbf{B} = 0$ in cylindrical coordinates. 

To establish reasonable values for the jet energy and power, we provide some estimates here. Assuming that the jet energy is extracted from the spin of the black hole, it can be estimated from the black hole's rotational energy \citep{bookMeier12}:
\begin{eqnarray} 
\rm E_{rot} = \left(1-\sqrt{ \frac{1+\sqrt{1-a^2} }{2}} \right) M_{\rm bh}c^2, \label{Erot}
\end{eqnarray}
where $\rm a$ is the black hole spin parameter, $\rm M_{bh}$ is the black hole mass, and $\rm c$ is the speed of light.  Using multiwavelength data for GRSs, \citet{dabhade20b}  indicate that the central black holes in GRSs have low spin ($\rm a<0.1$) and masses of $\rm 10^8-10^9 M_{\odot}$  for redshifts of $z<1$. The corresponding typical rotational energy is $\rm \lesssim 1.3 \times 10^{-3} M_{bh} c^2$. In this work, we mainly adopt a jet energy of $\rm 6.2 \times 10^{-4} M_{bh} c^2$.

The jet power can be related to the black hole accretion rate as
\begin{eqnarray} 
\rm P_{jet}= \eta_{jet} \dot{M}_{bh} c^2 ,
\end{eqnarray}
where $\rm \eta_{jet}$ is the jet efficiency. The Eddington luminosity can be expressed as 
\begin{eqnarray} 
\rm L_{Edd}= 0.1 \dot{M}_{Edd} c^2 ,
\end{eqnarray}
where $\rm L_{Edd}$ is the Eddington luminosity, which we adopt as $\rm 1.3\times 10^{38} erg\ s^{-1} (M_{bh} / M_{\odot})$, $\rm \dot{M}_{Edd}$ is the Eddington accretion rate, and $0.1$ represents the standard radiative efficiency. Thus the ratio of jet power to Eddington luminosity is
\begin{eqnarray} 
\rm \lambda_{jet} = \frac{P_{jet}}{L_{Edd}} = 10 \eta_{jet} \dot{m} ,
\end{eqnarray}
where $\rm \dot{m} = \dot{M}_{bh} / \dot{M}_{Edd} $ is the accretion rate in units of the Eddington accretion rate.  

  The ratio of jet power to Eddington luminosity $\rm \lambda_{jet}$ can be estimated from standard accretion physics and general relativistic MHD simulations (see \citet{yuan14} and \citet{davis20} for reviews). Jets can be launched when black holes are in sub-Eddington hot accretion or super-Eddington accretion states \citep{Tchekhovskoy15}. For a hot accretion flow, the AGN luminosity is typically below $\rm 0.02L_{Edd}$ \citep{yuan18} and the accretion rate is $\rm \dot{m} \lesssim 0.01$ \citep{bookMeier12, xie12, xie19}. For super-Eddington accretion, the accretion rate typically exceeds $\rm \dot{m} > 1$ \citep{bookMeier12}. For a thick, magnetically arrested accretion disk, the jet efficiency can be estimated as $\rm \eta_{jet} \lesssim 1.3 a^2$ \citep{Tchekhovskoy15, sadowski16, davis20}. Adopting a maximum jet efficiency of $1.3\%$ for low-spin black holes with $\rm a \lesssim 0.1$,  we obtain a jet-power-to-Eddington-luminosity ratio of $\rm \lambda_{jet} \lesssim 10^{-3}$ for hot accretion flows. In this work, we mainly adopt $\rm \lambda_{jet} \approx 5\times 10^{-4}$. Combined with a jet energy of $\rm 6.2 \times 10^{-4} M_{bh} c^2$, this power corresponds to a jet duration of about 548 Myr. We note that the typical lifetime of GRS jets is usually longer than that of radio lobes or X-ray cavities on smaller scales \citep{fabian12, dabhade23}, and may exceed a Gyr for GRSs of extreme size \citep{oei24a}. The values of jet energy and power in all our runs are listed in Table \ref{tab1}.  The typical jet power estimated from observations of GRSs lies in the range $10^{41}$--$10^{45}\ \mathrm{erg\ s^{-1}}$ \citep{dabhade23}, which can be compared with the powers we adopt.

\subsection{Calculation of radio luminosity}
\label{section2.4}

\begin{figure*}
\centering
\includegraphics[height=0.36\textheight]{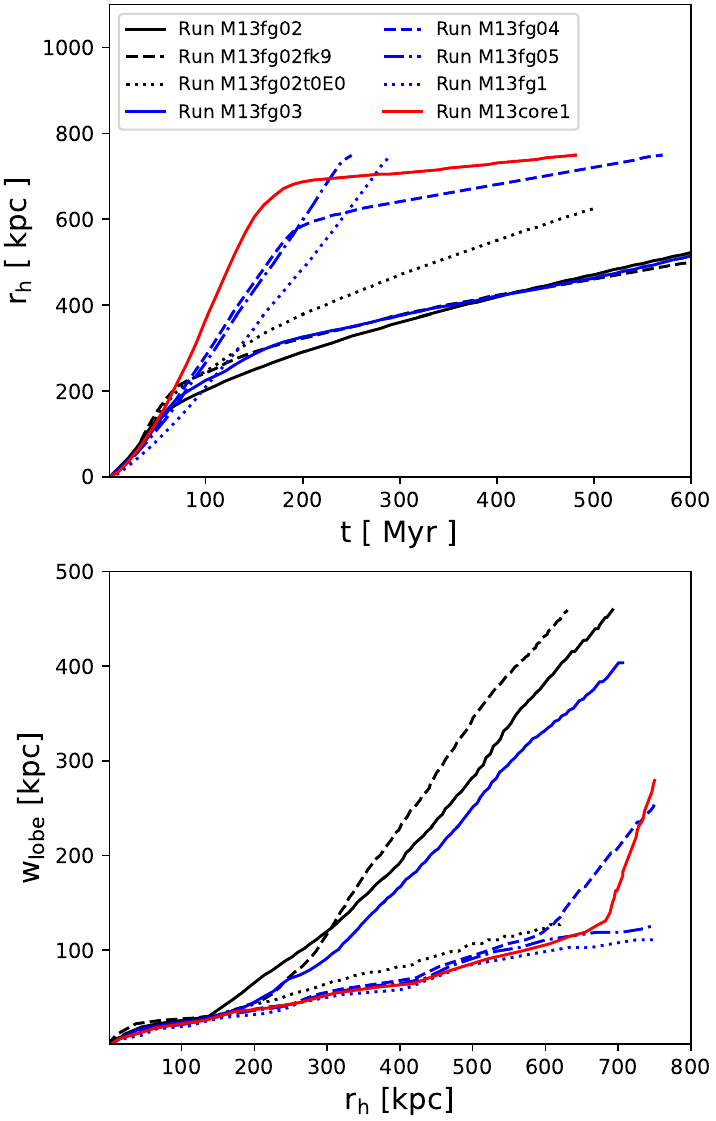}
\includegraphics[height=0.36\textheight]{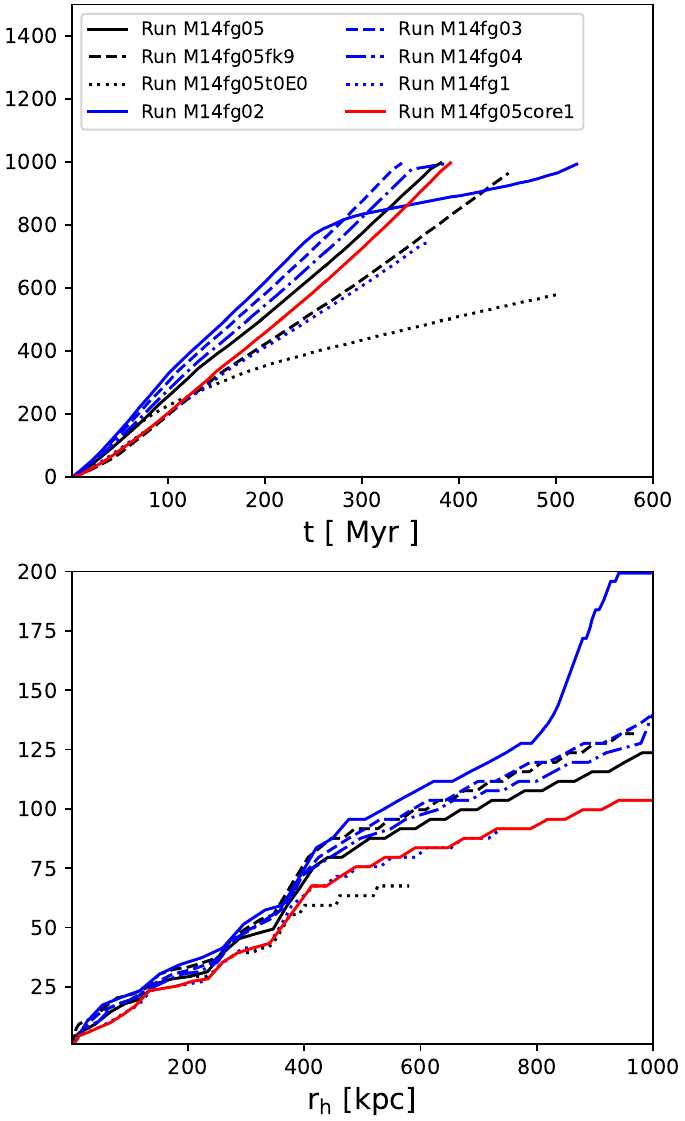}
\includegraphics[height=0.36\textheight]{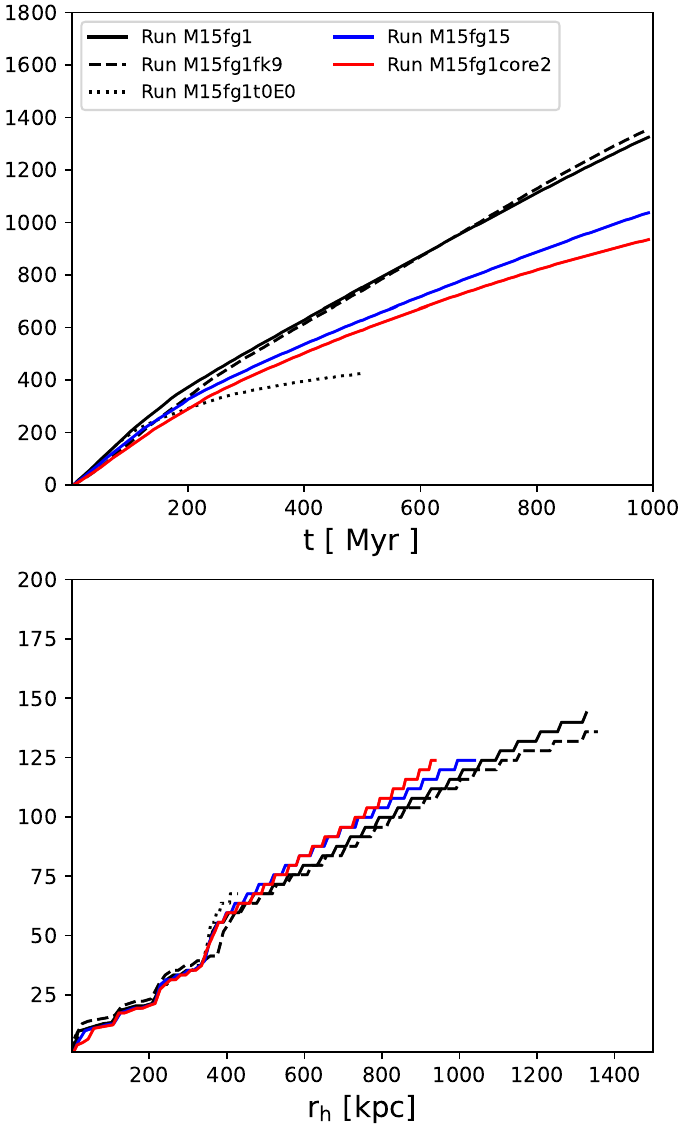}
\caption{  
 Evolution of the traveling distance $\rm r_h$ of the bubble heads (one-sided scale of the radio sources; top panels) and the maximum lobe width $\rm w_{lobe}$ as a function of $\rm r_h$ (bottom panels) in dark matter halos of $10^{13}$ (left), $10^{14}$ (middle), and $10^{15}$ (right) solar  masses.} 
\label{figrht-wr}
\end{figure*}

Similar to previous works, we calculate the radio emission using post-processing methods \citep{hardcastle14, yates18}.  The MHD simulation is single-fluid and does not specify the leptonic or hadronic composition. Our synchrotron calculation assumes a leptonic (electron) emitting population. On the other hand, ideal MHD does not include diffusive shock acceleration or reconnection-driven acceleration. Our post-processing therefore adopts a given non-thermal distribution rather than predicting the acceleration efficiency. We assume the distribution of non-thermal electrons within the radio sources follows
\begin{equation}  
\rm N(\gamma) = N_0 \gamma^{-q}, 
\label{N_electron}
\end{equation}
where $\rm \gamma$ is the Lorentz factor and $\rm q$ is the spectral index of the electrons. 
As an approximation, the distributions of electrons and radiation are typically assumed to be locally isotropic, allowing the emission coefficient to be expressed as
\begin{equation}  
\rm j_{\nu} = a(q) \frac{\sqrt{3}}{4\pi} \frac{e^3 N_{0}}{m_e c^2} \left(\frac{2\pi m_e c \nu}{3e}\right)^{-\frac{q-1}{2}} B^{\frac{q+1}{2}},
\end{equation}
with
\begin{equation}  
\rm a(q) = \frac{\sqrt{\pi}}{2} \Gamma\left(\frac{q+5}{4}\right) \left[\Gamma\left(\frac{q+7}{4}\right)\right]^{-1} f(q),
\end{equation}
 where $\rm \Gamma(y)$ is the $\Gamma$ function of argument y, and
\begin{equation}  
\rm f(q) = \frac{1}{q+1} \Gamma\left(\frac{q}{4} + \frac{19}{12}\right) \Gamma\left(\frac{q}{4} - \frac{1}{12}\right).
\end{equation}
The radiation exhibits a spectral distribution $\rm j_{\nu} \propto \nu^{-\alpha}$ with $\rm q = 2\alpha +1$.  \citet{hardcastle18} provided a semi-analytical model for emission from jetted lobes evolving to Mpc scales, including inverse-Compton losses from scattering of cosmic microwave background (CMB) photons. To approximately emulate the cooling of non-thermal electrons, we adopt a length-dependent spectral index  for radio emission $\rm \alpha$ for low-redshift radio sources following \cite{hardcastle18}:
\begin{equation}  
\rm \alpha \approx  0.1 [\lg(L/\text{kpc}) -1] +0.7,
\label{eqalpha}
\end{equation}
for $\rm 1\,\text{kpc} \lesssim L \lesssim 1\,\text{Mpc}$.
Given that the magnetic field is obtained from our simulations, the only unknown parameter is $\rm N_0$ in Eq. \ref{N_electron}. According to Eq. \ref{N_electron}, the energy density of non-thermal electrons is
\begin{equation}  
\rm e_e = \int_{\gamma_{\text{min}}}^{\gamma_{\text{max}}} N(\gamma) \gamma m_e c^2 \, d\gamma.  
\end{equation}
Thus, $N_0$ is determined by
\begin{eqnarray}  
\rm N_{0} = \left \{
\begin{aligned}
&\rm \frac{e_e}{m_ec^2} [\ln\left( \frac{\gamma_{max}}{\gamma_{min}}\right)]^{-1} &, \quad q=2 \\ \\
&\rm \frac{e_e}{m_ec^2}  \frac{2-q}{ \gamma_{max}^{(2-q)} - \gamma_{min}^{(2-q)}} &, \quad q\neq 2 \\
\end{aligned}
\right .
\label{N_0}
\end{eqnarray}
where $\gamma_{min}$ and $\gamma_{{max}}$ are empirically set to $100$ and $10^{5}$, respectively \citep{hardcastle14, hardcastle18}. Assuming equipartition between the non-thermal electron energy density and the magnetic energy density \citep{bookCondon16}, we have
\begin{equation} 
\rm e_e = \frac{4}{3} e_m, 
\label{e_e}
\end{equation}
with $\rm e_m$ the magnetic energy density. Finally, $\rm N_0$ can be derived from Eqs. \ref{N_0} and \ref{e_e}.

The radio power (luminosity) of the radio sources is  obtained by integrating over the radiative region as:
\begin{equation}  
\rm P_{\nu} = \int_V 4\pi j_{\nu} \, dV.
\end{equation}

 We clarify that the MHD simulation of jet evolution constitutes the main result of this work. The post-processing synchrotron calculation is intended as a supplementary diagnostic of how the simulated flow maps onto emission power. The calculation also cannot capture the coupled evolution of cooling and re-acceleration, the back-reaction of radiating particles on the flow, or additional high-energy channels. These limitations affect the quantitative interpretability of the supplementary emission power.

\begin{figure*}
\centering
\includegraphics[height=0.25\textheight]{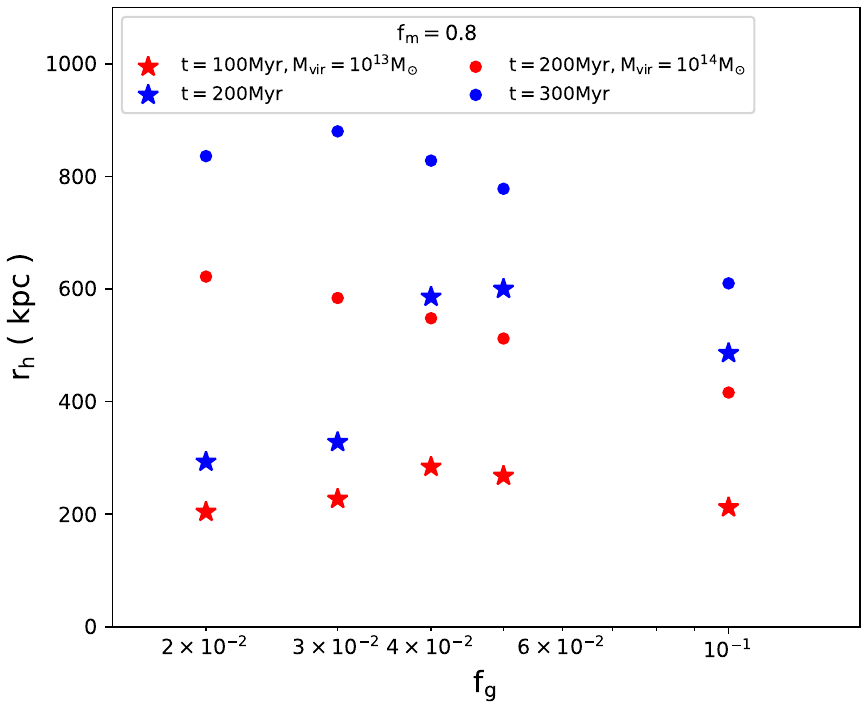}
\includegraphics[height=0.25\textheight]{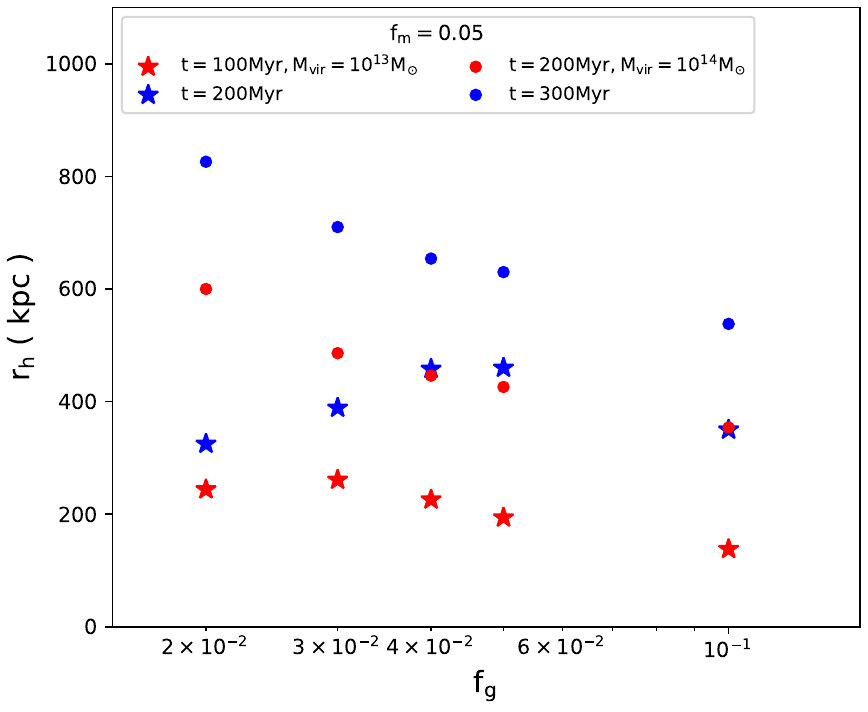}
\caption{ The traveling distance $\rm r_h$ of the bubble heads for magnetic (left, $\rm f_m=0.8$) and kinetic (right, $\rm f_m=0.05$) jets varying with hot gas fraction in dark matter halos of $10^{13}$ (stars) and $10^{14}$ (dots) solar masses.}   

\label{figrh-fg}
\end{figure*}

\begin{figure*}
\centering
\includegraphics[height=0.25\textheight]{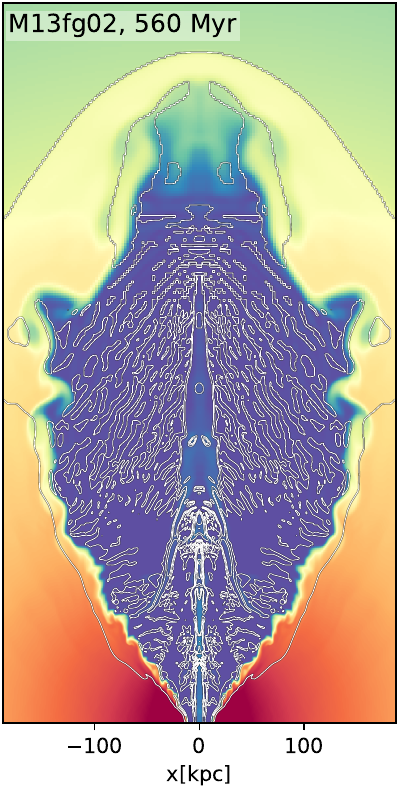}
\includegraphics[height=0.25\textheight]{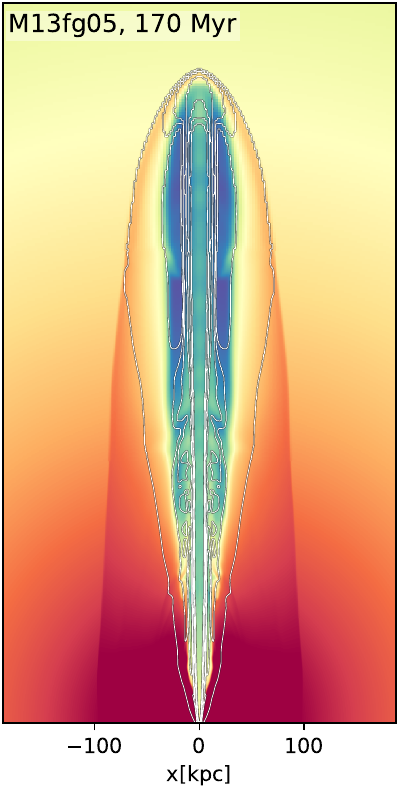}
\includegraphics[height=0.25\textheight]{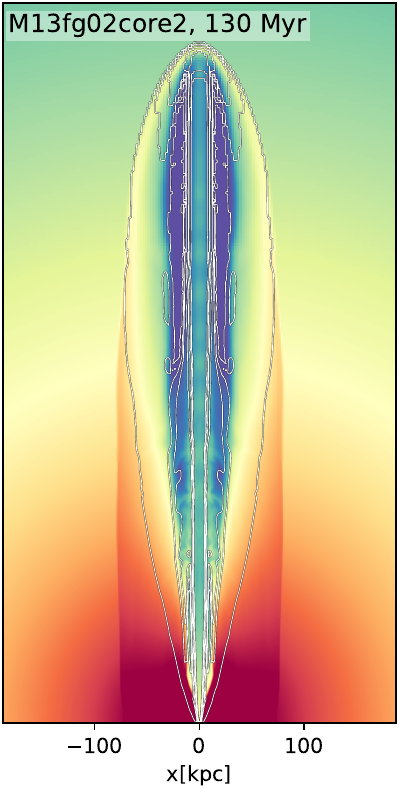}
\includegraphics[height=0.25\textheight]{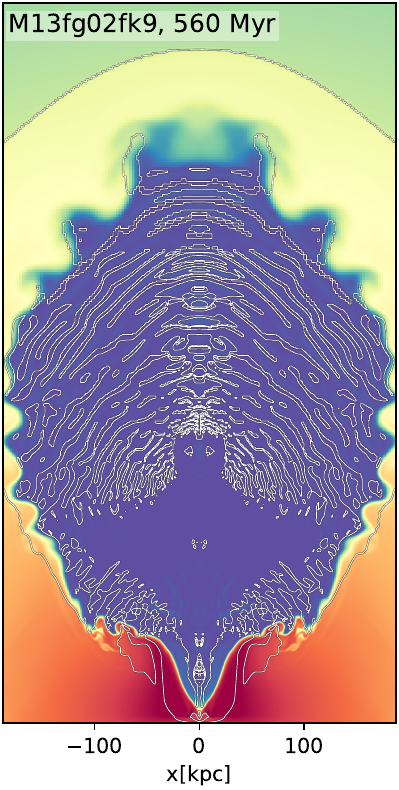}
\includegraphics[height=0.25\textheight]{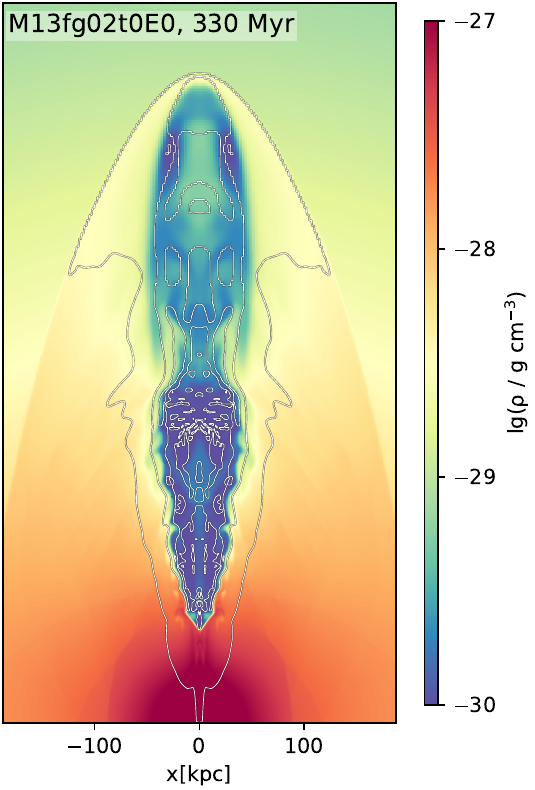}\\
\includegraphics[height=0.25\textheight]{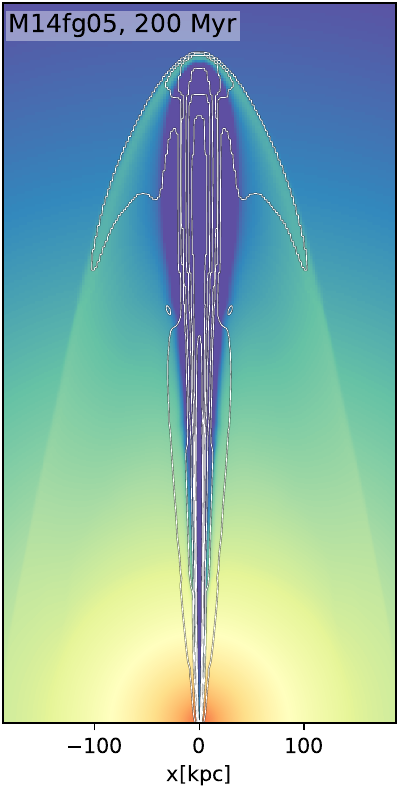}
\includegraphics[height=0.25\textheight]{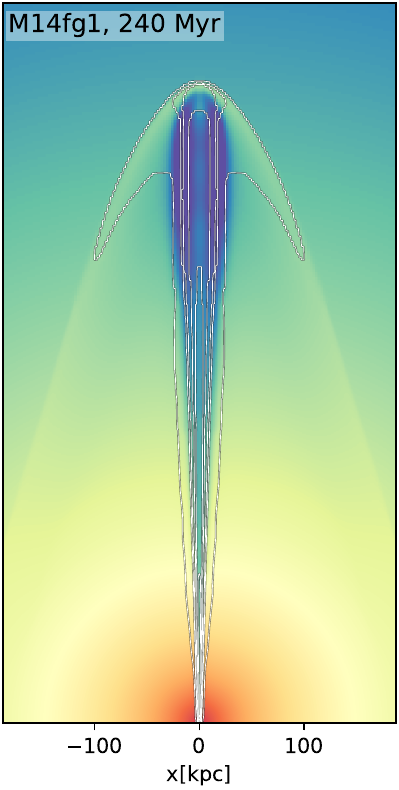}
\includegraphics[height=0.25\textheight]{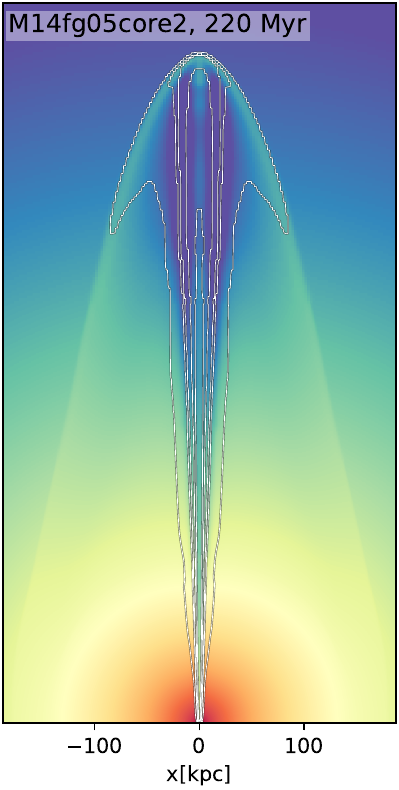}
\includegraphics[height=0.25\textheight]{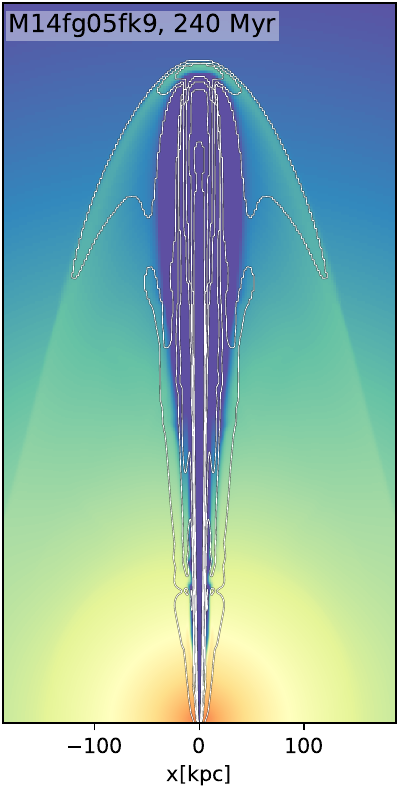}
\includegraphics[height=0.25\textheight]{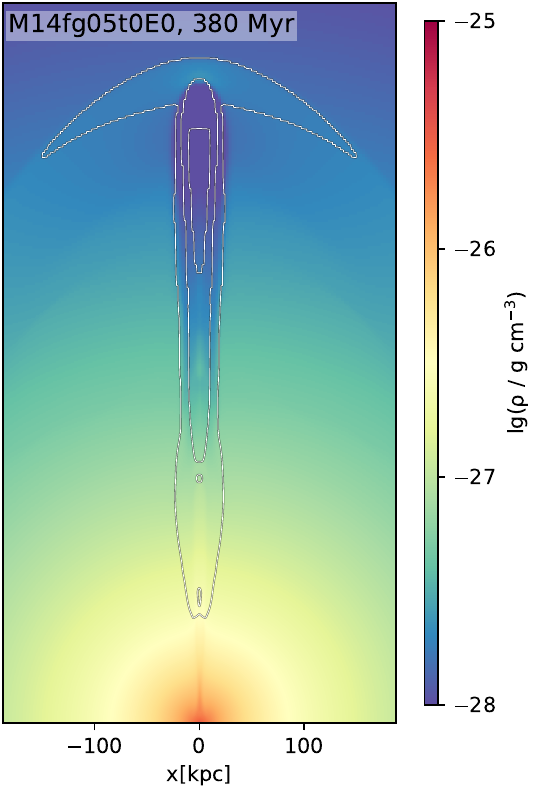}\\
\includegraphics[height=0.25\textheight]{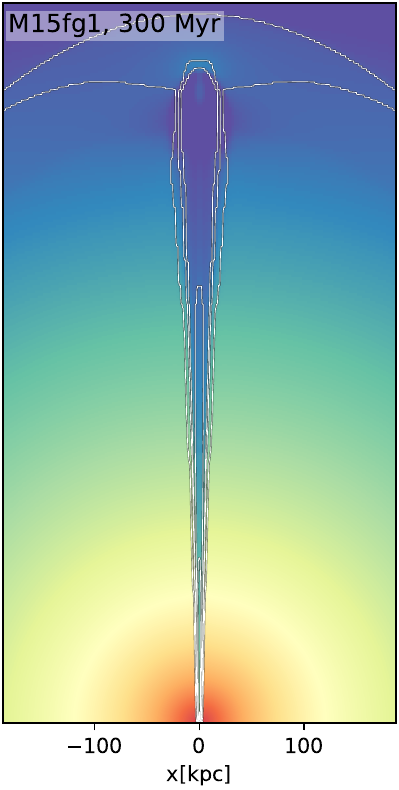}
\includegraphics[height=0.25\textheight]{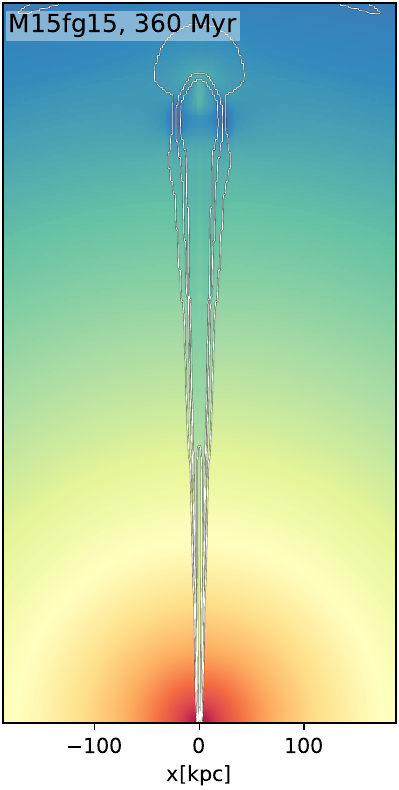}
\includegraphics[height=0.25\textheight]{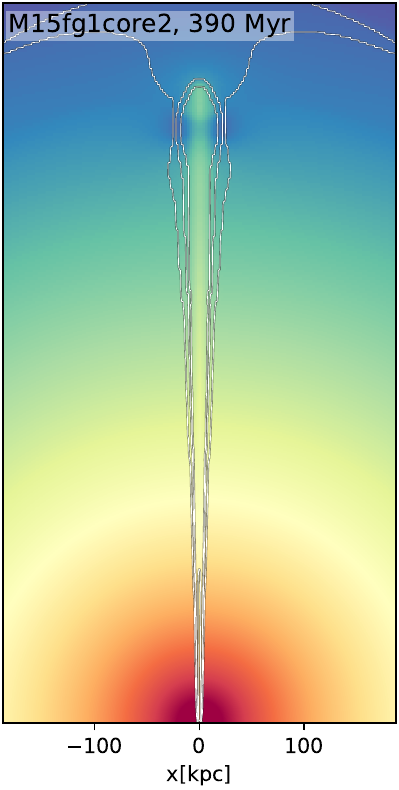}
\includegraphics[height=0.25\textheight]{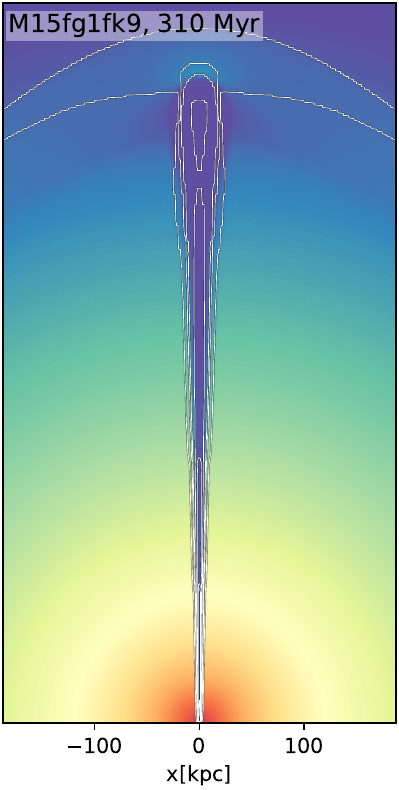}
\includegraphics[height=0.25\textheight]{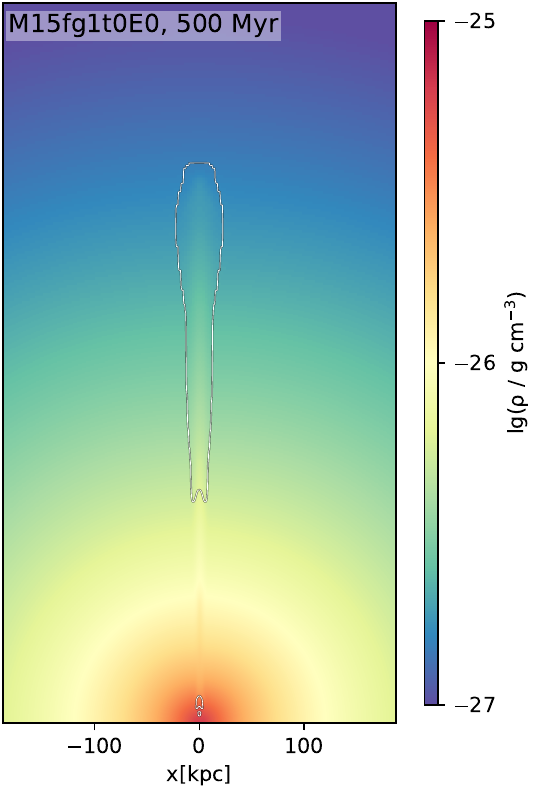}
\caption{Density snapshots of representative runs, taken when the bipolar radio lobes reach a scale of about 1 Mpc (one-sided scale $\rm r_h \sim 500$ kpc); the evolution of the scale of these runs is shown in Fig.~\ref{figrht-wr}. The three rows correspond to dark matter halos of $10^{13}$ (top), $10^{14}$ (middle), and $10^{15}$ (bottom) $\rm M_\odot$. The name of the run and the corresponding time are given at the top of each panel. The color encodes the base-10 logarithm of the gas density (in $\rm g\ cm^{-3}$), and the white lines show the contours of the vertical velocity ($\rm v_z$) at $\rm (0.05, 0.5, 1, 3, 5, 7, 9)\times 10^{8}\ cm\ s^{-1}$. The jetted lobes appear as low-density cavities: they are broad and nearly spherical in the low-mass halos with tenuous atmospheres, become narrower and more elongated as the central density increases, and detach from the halo center or are disrupted in the short-duration runs (right panels).}
\label{figden2d}
\end{figure*}

\section{Results and discussion}
\label{section3}                 
\subsection{Evolution of scale and morphology}
\label{section3.1}
\label{p-d}

\begin{figure*} 
\centering
\includegraphics[height=0.2\textheight]{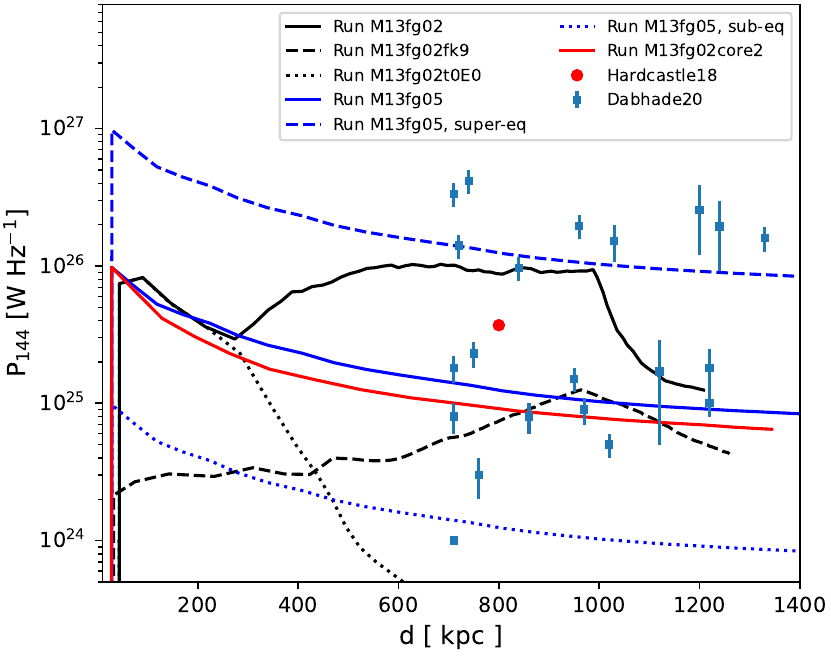}
\includegraphics[height=0.2\textheight]{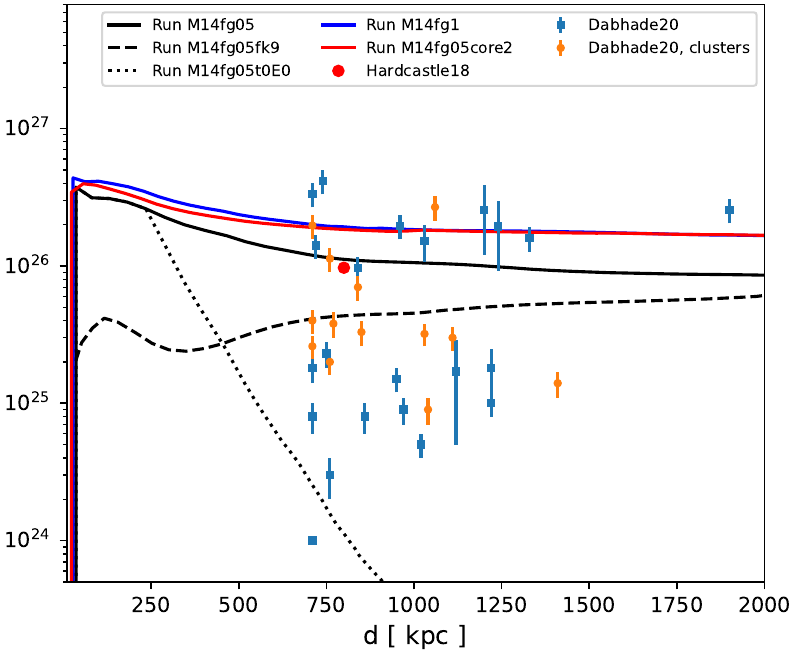}
\includegraphics[height=0.2\textheight]{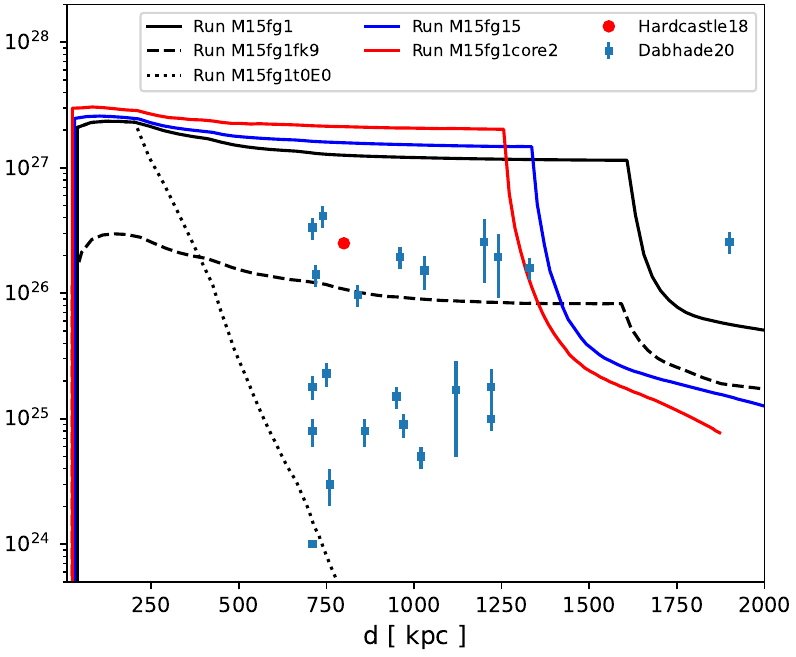}
\caption{Evolution of the simulated radio sources in the P-D diagram (from left to right for halo masses of $\rm 10^{13}$, $\rm 10^{14}$, and $\rm 10^{15} M_{\odot}$). The linear size ($\rm d = 2r_{h}$) denotes the two-sided scale of the simulated sources. The radio power is calculated at $144(1+z)$ MHz, assuming a redshift $\rm z=0.2$. The blue data points are taken from \cite{dabhade20a} for sources in the redshift range $\rm 0.1 < z < 0.3$, while the orange data points correspond to sources at the centers of identified clusters with halo masses of $0.7-2\times 10^{14} M_\odot$ and redshifts $\rm 0.1 < z < 0.4$. The red dots represent the values predicted by the linear relation of \cite{hardcastle18}.  'super-eq' and 'sub-eq' in the left panel denote cases with electron energies ten times larger and ten times smaller than the equipartition value (Eq.~\ref{e_e}).} 
\label{figPd}
\end{figure*}

\begin{figure*}
\centering
\includegraphics[height=0.2\textheight]{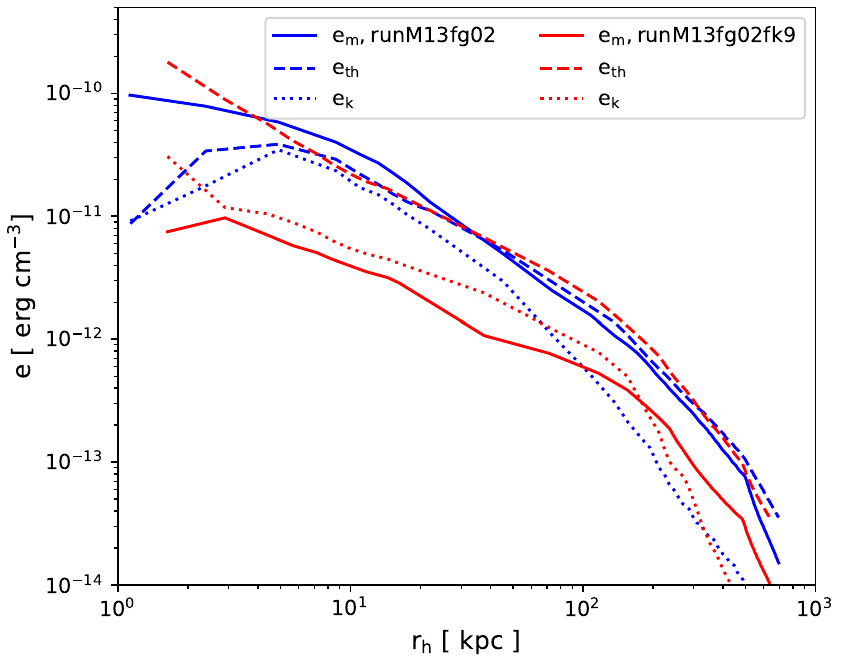}
\includegraphics[height=0.2\textheight]{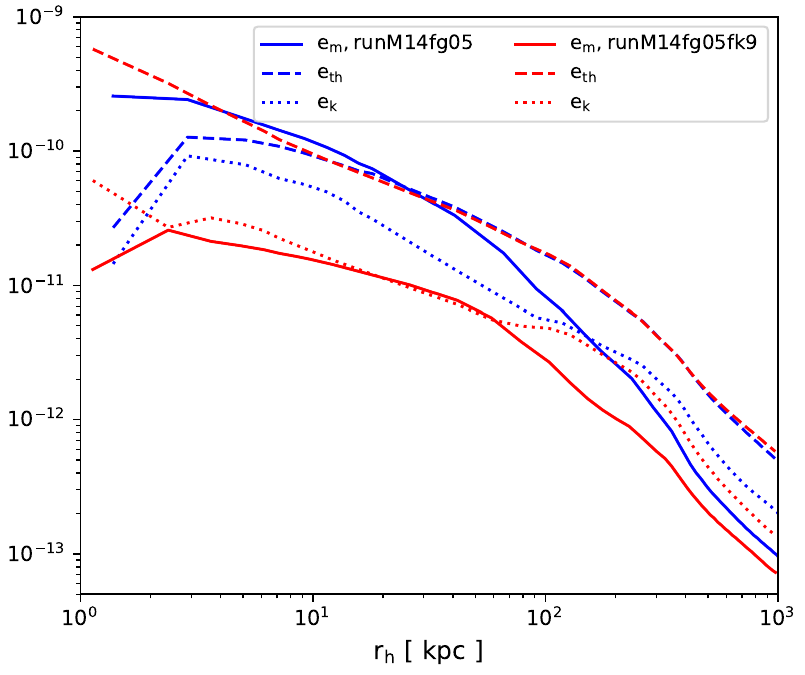}
\includegraphics[height=0.2\textheight]{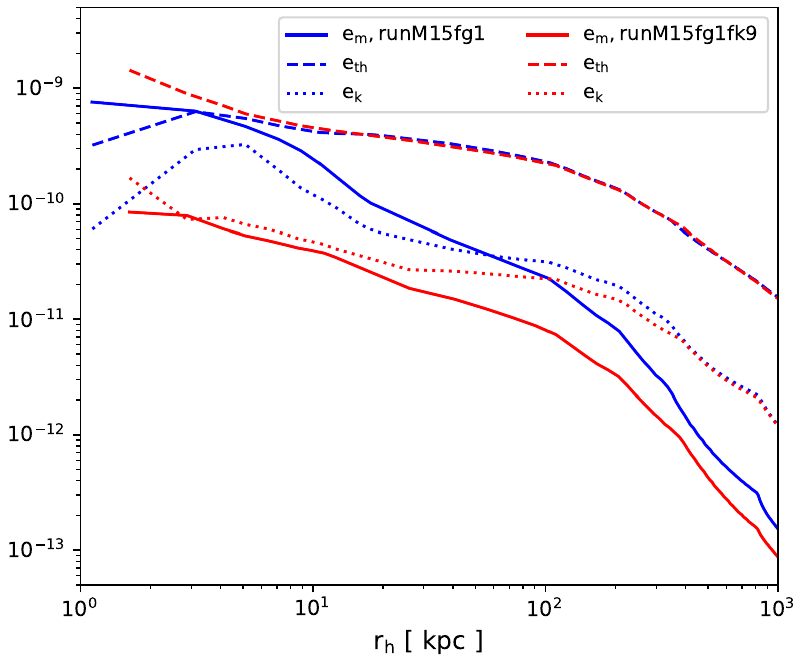}
\caption{Volume-averaged energy densities of different components as a function of lobe traveling distance $\rm r_h$ for runs with initially kinetic-energy-dominated jets and those with magnetic-energy-dominated jets. Different colors indicate different runs, while dotted, dashed, and solid lines represent kinetic, thermal, and magnetic energy densities ($\rm e_{k}$, $\rm e_{th}$, and$\rm e_m$), respectively.}
\label{figei}
\end{figure*}

The scale evolution of the radio sources in our runs is shown in the top panels of Fig. \ref{figrht-wr} as the traveling distance $\rm r_h$ of the jetted bubble heads (one-sided scale of the simulated sources) varying with time. Since the magnetic field in the hot gas atmosphere is negligible compared to that in the low-density jetted bubbles in our simulations (Sect. \ref{section2.3}), the radio lobe (jetted bubble) region is identified by a density below a threshold and a toroidal magnetic field stronger than $10^{-9} \rm{G}$. The density threshold is set to $5\times 10^{-27} \rm{g\ cm^{-3}}$ for runs with $\rm 10^{13}$ and $\rm 10^{14}$ solar mass halos, and to $5\times 10^{-25} \rm{g\ cm^{-3}}$ for the $\rm 10^{15}$ solar mass halo run. The accuracy of this method has been verified by comparing the results in Fig. \ref{figrht-wr} with the density snapshots in Fig. \ref{figden2d}. As shown in Fig. \ref{figrht-wr}, most radio sources grow to become GRSs ($\rm 2r_h > 0.7$ Mpc). Given the normal values adopted for the hot gas fraction and the reasonable jet energy and power (Sects. \ref{section2.2} and \ref{section2.3}), the formation of GRSs in these runs indicates that an unusually low-density gas environment is not a prerequisite for GRS formation.

The propagation of radio lobes is significantly influenced by the  gas environment within the dark matter halos, particularly for sources in dark matter halos of mass $\rm 10^{13}\ M_{\odot}$ (left top panel in Fig. \ref{figrht-wr}). When the central gas density is increased by adopting a higher hot gas fraction ($\rm f_g=0.05$ in run  M13fg05) or a lower core parameter ($\rm a_{core}=0.2$ in run  M13fg02core2), the radio lobes can propagate faster than in the fiducial run  ($\rm f_g=0.02,\ a_{core} =0.5$ in run  M13fg02). This likely occurs because a higher-pressure atmosphere collimates the jet more effectively, making it more penetrating than in a rarefied environment and thus promoting jet propagation. This is also reflected in the slimmer jetted lobes in denser environments (left bottom panel of Fig. \ref{figrht-wr}). By changing the density profile of the isothermal beta atmosphere, \cite{english19} found a similar phenomenon. On the other hand, a higher-pressure atmosphere can also enhance the interaction between the jet and the surrounding medium, which tends to suppress propagation. When the central density is further increased in a supplemental run ($\rm f_g=0.1$ in run  M13fg1), this promotional effect is diminished. In halos of mass $\rm 10^{14}\ M_{\odot}$ and $\rm 10^{15}\ M_{\odot}$, increasing the central gas density even weakens jet propagation to some degree compared with the fiducial run.   The traveling distances $\rm r_h$ of the bubble heads as functions of hot gas fraction in dark matter halos of $10^{13}$ and $10^{14}$ solar masses are shown in Fig.~\ref{figrh-fg}. We can see that the relation between $\rm r_h$ and $\rm f_g$ is not simply linear in halos of $\rm 10^{13}\ M_{\odot}$. These results reveal a much more complicated role of the gas environment in jet propagation than the previous suggestion that a low-density environment promotes GRS formation.  

The morphologies of the jetted lobes differ significantly among the runs. The maximum lobe widths as a function of length are shown in the bottom panels of Fig. \ref{figrht-wr}. Snapshots of the density distribution in our runs, captured when the bipolar radio lobes reach a scale of about 1 Mpc (one-sided scale $\rm r_h \sim 500$ kpc), are shown in Fig. \ref{figden2d}. As can be seen, the jetted lobes in runs  M13fg02 and M13fg02fk9 are much wider than those in the other runs. This suggests that wider GRSs form more readily from hosts in lower-mass dark matter halos with low-density gas cores, where they undergo greater adiabatic cooling in the lower-pressure environment. 

The dependence of lobe morphology on jet composition and duration is also explored. As shown in Figs. \ref{figrht-wr} and \ref{figden2d}, the morphology of jets initially dominated by kinetic energy (run names containing "fk9") is more similar to that of the fiducial runs ( M13fg02, M14fg05, and M15fg1, initially dominated by magnetic energy) than to runs with different gas environments. Runs with shorter jet durations (54.8 Myr; run names containing ``t0E0'') can detach from the halo center (run  M13fg02t0E0) or even become severely disrupted by the dense environment (run  M15fg1t0E0) by the time they reach GRS scales (see Fig. \ref{figden2d}). This indicates that low-power jets (as in this work) with short active durations (tens of Myr) can only appear as remnant sources when they propagate to GRS scales. Indeed, such sources rapidly become faint at large scales (Sect. \ref{p-d}). 

\subsection{P-D diagram}
\label{section3.2}

The P-D diagram is an important diagnostic for studying radio sources with different jet powers \citep{baldwin82,hardcastle18}. In this work, we calculate the radio power (or luminosity) using the method updated from \cite{duan25} (Sect. \ref{section2.4}). The identification of the radiative region follows the same approach used for measuring the scale of the simulated sources (Sect. \ref{section3.1}). The evolution of the GRSs  in the P--D diagram is shown in Fig. \ref{figPd}, where the linear size ($\rm d = 2r_{h}$) denotes the two-sided scale of the simulated sources.  \citet{dabhade20a} matched the GRSs with host halo masses around $\rm 10^{14} M_{\odot}$. The simulated sources are placed at redshift $z=0.2$, corresponding to the redshift at which the hot gas fractions we use were measured \citep{dev24}. The radio power is calculated at a frequency of $(z+1)144$ MHz. Observed GRSs without matched host halo masses, selected within the redshift range $\rm 0.1 < z < 0.3$, are shown as blue data points in Fig. \ref{figPd}. The orange data points in the middle panel of Fig. \ref{figPd} correspond to sources at the centers of identified clusters with halo masses in the range $0.7-2\times 10^{14} M_\odot$ and redshifts $\rm 0.1 < z < 0.4$.

As shown, the radio powers of  most of the simulated GRSs within halo masses of $\rm 10^{13}$ and $\rm 10^{14} M_\odot$ are comparable to those observed at similar scales. Comparing the radio powers of the jetted lobes in  halos of different masses, the radio power of GRSs generally increases with dark matter halo mass (and consequently with jet energy and power in our simulations). The simulated sources with the same power but shorter jet duration ($\rm t_{jet} = 54.8$ Myr; run names containing `t0E0') fade much more rapidly than those in other runs. Indeed, these sources with low power ($\rm \sim 10^{-4} L_{Edd}$) and short jet duration (tens of Myr) have become faint remnant sources by the time they reach GRS scales.

\begin{figure*}
\centering
\includegraphics[height=0.4\textheight]{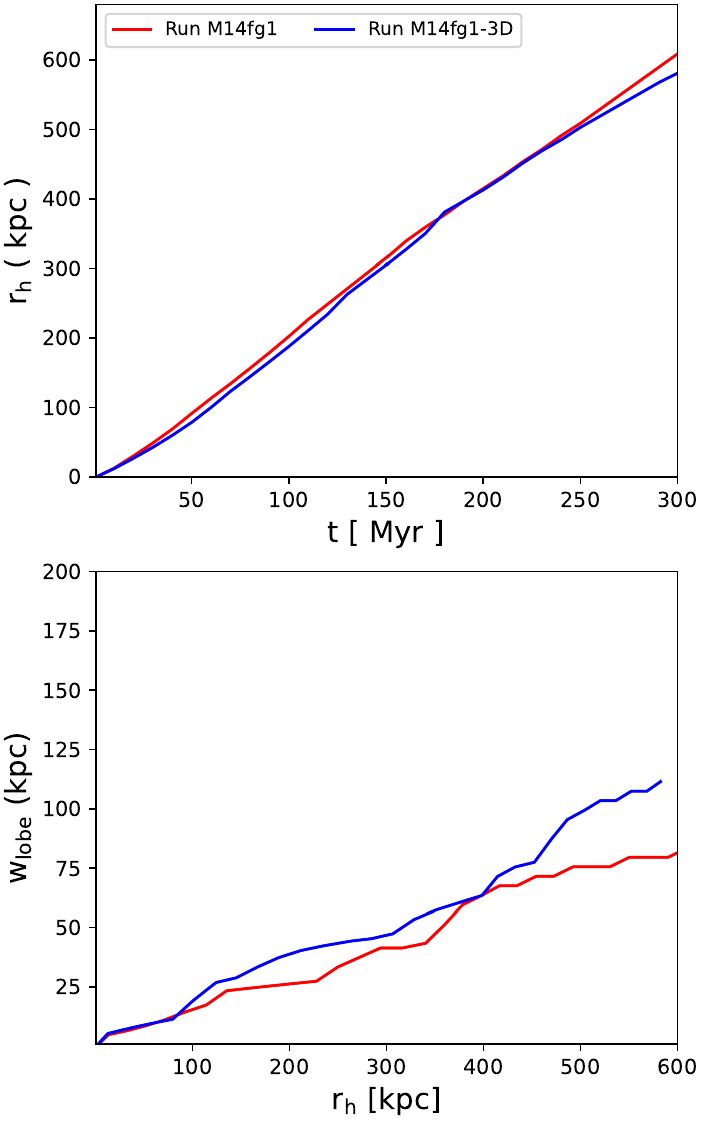}
\includegraphics[height=0.4\textheight]{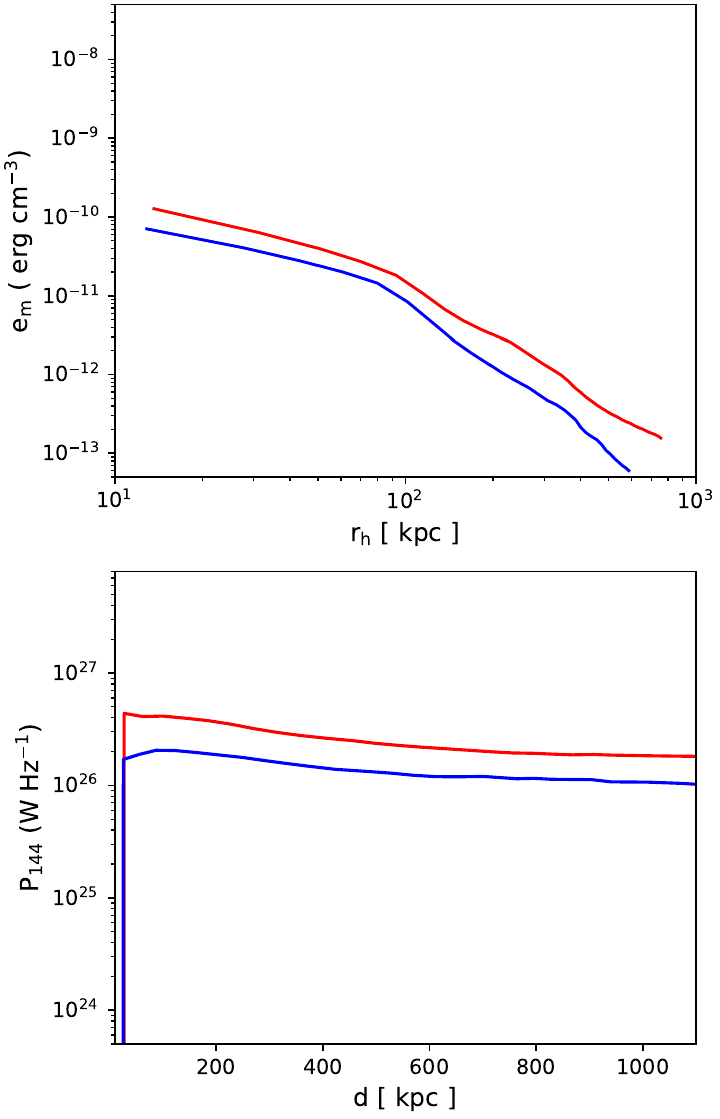}
\caption{  
 Comparison between the two-dimensional axisymmetric (2.5D) run M14fg1 (red lines) and the full three-dimensional run M14fg1-3D (blue lines). Left: evolution of the traveling distance $\rm r_h$ of the bubble heads (top) and the maximum lobe width $\rm w_{lobe}$ as a function of $\rm r_h$ (bottom). Right: volume-averaged magnetic energy densities of the different components as a function of lobe traveling distance $\rm r_h$ (top), together with the P--D diagram of the two runs (bottom, similar to Fig.~\ref{figPd}).}
\label{figrht-2Dvs3D}
\end{figure*}

The linear relation between jet power and radio power is widely used in the literature \citep{hardcastle18,dabhade20b,ye25}. By applying a semi-analytical model informed by simulations and observations, the linear relation for active radio sources at low redshift is given by \citet{hardcastle18}:
\begin{equation}  
\rm P_{150\,MHz} = 3\times 10^{27} (P_{jet}/10^{38}\ W)\ W\ Hz^{-1}.
\end{equation}
Using this relation and considering a spectral distribution $P_{\nu} \propto \nu^{-\alpha}$ with $\alpha \sim 0.86$ (Eq. \ref{eqalpha}) for a lobe length of 400 kpc (d=800 kpc), the radio powers corresponding to the jet powers adopted in this work (listed in Tab. \ref{tab1}) are shown as red dots in Fig. \ref{figPd}. These values are typically higher than the radio powers of jets injected with kinetic energy dominance (black dashed lines) in our simulations  (the jet model in \cite{hardcastle18} is kinetic-energy-dominated). This result can be roughly understood from the different assumptions regarding the electron energy density in the radiation model. We have assumed equipartition between electron energy and magnetic energy when calculating the radiation, whereas in \cite{hardcastle18} the electron energy is set to be ten times higher than the magnetic energy.  Consequently, the radiation from the sources in \cite{hardcastle18} is stronger than that of our kinetic jets (black dashed lines in Fig. \ref{figPd}).

\subsection{Uncertainties and limitations}
\label{section3.3}

The composition of active galactic nucleus jets remains an unsolved problem \citep{hardcastle20}, and the accretion state of the central black holes in GRS host galaxies is also unclear. Furthermore, different jet formation models (e.g., the Blandford-Znajek, Blandford-Payne, and magnetic tower models) predict different jet compositions \citep{yuan14}. Therefore, careful consideration of jet parameters is essential when simulating jet feedback processes. 
On the other hand, \cite{gan17} found that a head–tail radio morphology is well reproduced by using a magnetic energy dominant jet model, whereas lobes of jets injected with kinetic or thermal energy dominance could not survive the disruptive effects of intracluster weather. \cite{duan25} shows that jets injected with toroidal magnetic field energy dominance tend to produce larger radio sources than those dominated by kinetic or thermal energy. Based on these findings, we have chosen in this work to simulate jets with magnetic field energy as the dominant  component, while jets injected with kinetic energy dominance are also simulated for parameter studies.  The propagation of the kinetic jets is affected similarly by the hot baryonic gas fractions in the dark matter halos, as shown in the right panel of Fig.~\ref{figrh-fg}. We also note that even when jets are initially injected with magnetically dominated energy, their lobes become dominated by kinetic or thermal energy as they evolve to larger scales from the host galaxy center,  so that most of the volume of a GRS-scale lobe is not magnetically dominated, as shown in Fig. \ref{figei}. 

The relation between electron energy and magnetic energy is crucial in our post-processing radiative model (Sect. \ref{section2.4}). We have adopted the equipartition assumption (Eq. \ref{e_e}), which is widely used in radio astronomy \citep{bookCondon16, dabhade23, wu25} but whose validity remains under debate \citep{hardcastle18, hardcastle20}.  As the radiation power is proportional to the electron energy, a super-equipartition or sub-equipartition electron energy density will make the radiation power higher or lower (blue dashed and dotted lines in the left panel of Fig.~\ref{figPd}). The ratio of electron energy density to magnetic energy density varies over a broad range (roughly from 1 to 100) in observations \citep{croston05, isobe15}. In this respect, our radiative model provides a lower limit for the radio luminosity of the simulated sources. On the other hand, although the magnetic field energy (and  also the electron energy) measured in observations is typically weaker for larger radio sources, consistent with our simulations (Fig. \ref{figei}), there remains a substantial spread (nearly two orders of magnitude) for sources of the same scale \citep{isobe15, dabhade23}. The difference in magnetic field energy density at large scales (e.g. $ \sim 1$ Mpc) between simulated sources with initially kinetic-energy-dominated jets and those with magnetic-energy-dominated jets falls within one order of magnitude and lies well within the observed range (for 1 Mpc sources, $\rm e_{B} \sim 10^{-15} - 10^{-11} erg\ cm^{-3}$ and $\rm e_{e} \sim 10^{-13} - 10^{-11} erg\ cm^{-3}$). Thus, it is unfeasible to constrain the jet energy composition by comparing the radio luminosities from MHD simulations with observations. 

\begin{figure}[!t] 
\centering
\includegraphics[height=0.40\textheight]{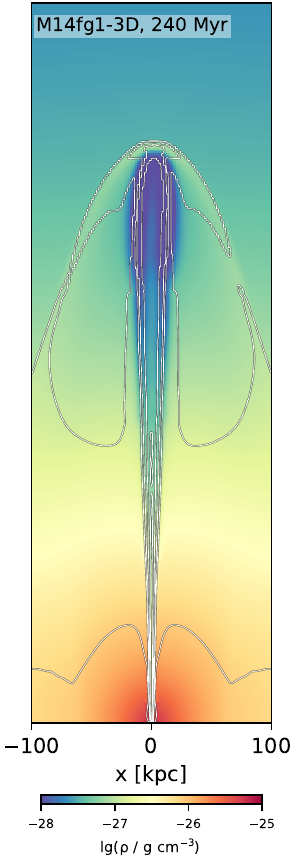}
\includegraphics[height=0.40\textheight]{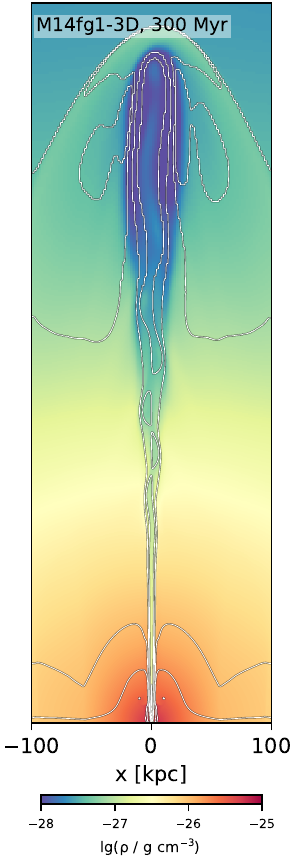}
\caption{Slices of the density distribution in the full three-dimensional run M14fg1-3D at $\rm t=240$ Myr (left) and $\rm 300$ Myr (right), similar to Fig.~\ref{figden2d}, showing the non-axisymmetric structures produced by the growth of the kink instability.}
\label{figden-3D}
\end{figure}

Another factor influencing the properties of simulated GRSs is the dimensionality of the simulations, as discussed in \cite{duan25}. For jets injected with magnetically dominated energy, the two-dimensional axisymmetric  (2.5D) MHD setup used here can suppress certain MHD instabilities—notably the kink instability—that would develop in full three-dimensional (3D) simulations \citep{mignone10}. The kink instability triggers non-axisymmetric features such as jet wiggling, which cannot be captured in our axisymmetric simulations. The dimensionality also makes a difference for jetted bubble collimation and for whether they evolve into Fanaroff--Riley type I or II sources \citep{perucho10, massaglia16}. However, our focus is not on the detailed morphology of the jet lobes but primarily on their final spatial extent. Importantly, \cite{perucho19} found that hydrodynamic jet propagation in 3D simulations can be faster than in 2D axisymmetric simulations, until the dentist drill effect produced by the growth of helical instabilities slows the 3D jet head, and the overall lobe scale remains similar in axisymmetric and three-dimensional MHD simulations, as demonstrated by \cite{mignone10}.  The comparison of the evolution of the traveling distance $\rm r_h$ of the bubble heads in runs M14fg1 (2.5D) and M14fg1-3D is shown in the top left panel of Fig.~\ref{figrht-2Dvs3D}. We can see that on the scales of primary interest ($\rm 2r_h \sim 1\ Mpc$), three-dimensional effects do not significantly influence the scale evolution of the lobes; three-dimensional effects such as the kink instability become obvious at later stages of the evolution, as shown in Fig.~\ref{figden-3D}. On the other hand, we can also see that 3D kink instabilities result in a larger maximum lobe width, a lower volume-averaged magnetic energy density, and consequently a synthetic synchrotron luminosity lower by a factor of about 1.5--2, as shown in Fig.~\ref{figrht-2Dvs3D}.

\section{Conclusions}
\label{section4}

We use MHD simulations to study the formation of GRSs from central galaxies in dark matter halos of $10^{13}$, $10^{14}$, and $10^{15}$ $\rm M_{\odot}$, adopting  hot gas fractions in the ranges 0.02--0.1, 0.02--0.1, and 0.1--0.15. The main results are:

(i) GRSs form in all three halo masses  with normal hot gas fractions when jets are injected with an energy of $\sim0.06\%$ of the central black hole's relativistic energy and a power of $\sim0.05\%$ of the Eddington luminosity. 

(ii) The dependence of the lobe propagation on the gas environment is non-monotonic. The propagation of radio lobes can be slower in halos with both sufficiently low and high central densities. By comparing simulated sources in $\rm 10^{13}\ M_{\odot}$ halos with low ( e.g., $\rm f_g=0.02,\ a_{core}=0.5$) and high central density ( e.g., $\rm f_g=0.05$ or $\rm a_{core}=0.2$), we find that lobe propagation can be slower in low-density halos because the lower central pressure fails to sufficiently collimate the jet, producing wider lobes that are less penetrating than those in denser atmospheres. Conversely, an atmosphere with sufficiently high pressure (e.g., in the $10^{14}$ and $10^{15}$ $\rm M_{\odot}$ halos studied here) can also enhance jet--medium interactions, thereby suppressing jet propagation. 

(iii) Assuming equipartition between non-thermal electron and magnetic energy, we have shown the P-D diagram of the simulated radio sources. Most of the simulated sources within halo masses of $\rm 10^{13}$ and $\rm 10^{14} M_\odot$ can reach radio powers comparable to observed GRSs at similar sizes, with radio power generally increasing with dark matter halo mass. The simulated sources with the same low power (0.05 percent of the Eddington luminosity) but a shorter jet duration ($54.8$ Myr) than other sources become faint remnant sources by the time they reach GRS scales.  However, we clarify that the post-processing synchrotron calculation is intended only as a supplementary diagnostic of how the simulated flow maps onto emission power. It cannot capture the coupled evolution of cooling and re-acceleration processes or additional high-energy channels; these limitations affect the quantitative interpretability of the supplementary emission power.


\begin{acknowledgements}
 X.D. thanks the anonymous referee for the comprehensive reports. This work was supported by the Natural Science Foundation of Henan Province (No.262300422635) and the High Performance Computing Center of Henan Normal University. AI (DeepSeek-V4.1) was used to perform grammar checks on the manuscript, and the author made selective revisions to the manuscript based on AI’s feedback.

\end{acknowledgements}


\bibliographystyle{apsrev4-2}
\bibliography{main}

\end{document}